\documentclass[jasa,12pt]{article}
\usepackage{mathrsfs}
\usepackage{amsfonts,amsmath,latexsym,amsthm}
\usepackage{graphicx} 
\usepackage{url, hyperref}
\usepackage{setspace}
\usepackage{natbib}
\bibpunct{(}{)}{;}{a}{}{,}
\usepackage{multirow}  
\usepackage{rotating}  
\usepackage{pst-all}      
\usepackage{pst-poly}     
\usepackage{multido}      
\usepackage{bm}
\usepackage{enumitem}

\usepackage{setspace}

\newtheorem{theorem}{Theorem}
\newtheorem{corollary}{Corollary}
\newtheorem{definition}{Definition}
\newtheorem{remark}{Remark}

\makeatletter
\newcommand{\specificthanks}[1]{\@fnsymbol{#1}}
\makeatother

\title{On the Use of Cauchy Prior Distributions for Bayesian Logistic Regression}
\date{}
\author{Joyee Ghosh\thanks{The University of Iowa, Iowa City, IA. 
       Email: {\tt joyee-ghosh@uiowa.edu}} \textsuperscript{,\specificthanks{3}}
       \and
		Yingbo Li\thanks{Clemson University, Clemson, SC. 
       Email: {\tt ybli@clemson.edu}} \textsuperscript{,}\thanks{These authors contributed equally.}
       \and
      		Robin Mitra\thanks{University of Southampton, Southampton, UK. 
       Email: {\tt R.Mitra@soton.ac.uk} }
 }

\begin{document}
\maketitle
\begin{abstract}

In logistic regression, separation occurs when a linear combination of 
the predictors can perfectly classify part or all of the observations in the sample, 
and as a result, finite maximum likelihood estimates
of the regression coefficients do not exist. 
\citet{Gelman_etal_2008} recommended
independent Cauchy distributions as 
default priors for the regression coefficients in logistic regression, 
even in the case of separation, and reported posterior modes in their analyses.
As the mean does not exist for the Cauchy prior, 
a natural question is whether the posterior means of the 
regression coefficients exist under separation. 
We prove theorems that provide necessary and 
sufficient conditions for the existence of posterior means
under independent Cauchy priors for the logit link and 
a general family of link functions, including the probit link.
We also study the existence of posterior means under
multivariate Cauchy priors.
For full Bayesian inference, we develop a Gibbs sampler 
based on P\'olya-Gamma data augmentation
to sample from the posterior distribution 
under independent Student-$t$ priors 
including Cauchy priors,
and provide a companion R package in the supplement.  
We demonstrate empirically that even when the posterior 
means of the regression coefficients exist under separation, 
the magnitude of the posterior samples for Cauchy priors 
may be unusually large, and the corresponding 
Gibbs sampler shows extremely slow mixing. 
While alternative algorithms such as the 
No-U-Turn Sampler in Stan can greatly improve mixing,
in order to resolve the 
issue of extremely heavy tailed posteriors for Cauchy priors under separation, 
one would need to consider lighter tailed priors such as normal priors
or Student-$t$ priors with degrees of freedom larger than one.

\end{abstract}



\doublespacing

\section{Introduction}

In Bayesian linear regression, 
the choice of prior distribution for the regression 
coefficients is a key component of the analysis. Noninformative priors are 
convenient when the analyst does not have much prior information, 
but these prior distributions are often improper which can lead to 
improper posterior distributions in certain situations.  
\citet{Fernandez_Steel_2000} investigated the propriety
of the posterior distribution and the existence of posterior moments
of regression 
and scale parameters for a linear regression model,
with errors distributed as scale mixtures of normals, under the
independence Jeffreys prior. For a design matrix of full column rank, 
they showed that posterior propriety holds under mild conditions on the sample size;
however, the existence of posterior moments is affected by the design matrix and the
mixing distribution.
Further, there is not always a unique choice of noninformative prior
\citep{Yang_Berger_1996}. On the other hand, proper prior distributions 
for the regression coefficients guarantee the propriety of 
posterior distributions. 
Among them, normal priors are commonly used in normal
linear regression models, as conjugacy permits efficient 
posterior computation. The normal priors are informative because
the prior mean and covariance can be specified 
to reflect the analyst's prior information, 
and the posterior mean of the regression coefficients is 
the weighted average of the maximum likelihood estimator
and the prior mean, with the weight on the latter decreasing
as the prior variance increases.

A natural alternative to the normal prior is the Student-$t$ prior distribution, 
which can be viewed as a scale mixture of normals.
The Student-$t$ prior has tails heavier than the normal prior,
and hence is more appealing in the case where weakly informative
priors are desirable. The Student-$t$ prior is considered robust,
because when it is used for location parameters, 
outliers have vanishing influence on 
posterior distributions \citep{Dawid_1973}.
The Cauchy distribution is a special case of the Student-$t$ distribution
with 1 degree of freedom. It has
been recommended as a prior for normal mean parameters in
a point null hypothesis testing \citep{Jeffreys_1961}, because
if the observations are overwhelmingly far from zero (the
 value of the mean specified under the point null hypothesis), 
the Bayes factor favoring the alternative hypothesis tends to infinity.
Multivariate Cauchy priors have also been proposed for regression coefficients
\citep{Zellner_Siow_1980}.

While the choice of prior distributions has been extensively studied for
normal linear regression, there has been comparatively less work for 
generalized linear models.  Propriety of the posterior distribution
 and the existence of posterior moments for binary response models 
under different noninformative prior choices have been considered 
\citep{Ibrahim_Laud_1991,Chen_Shao_2001}.

 Regression models for binary response variables may suffer from 
a particular problem known as separation, which is the focus of this paper.
For example, complete separation occurs if there exists a linear function of the covariates 
for which positive values of the function correspond to those units with 
response values of 1, while negative values of the function correspond 
to units with response values of 0. Formal definitions of separation \citep{Albert_Anderson_1984}, 
including complete separation and its closely related counterpart
quasicomplete separation, are reviewed in Section 2. 
Separation is not a rare problem in practice, 
and has the potential to become increasingly common
in the era of big data, with analysis often being made on data 
with a modest sample size but a large number of covariates. 
When separation is present in the data, \citet{Albert_Anderson_1984} 
showed that the maximum likelihood estimates (MLEs) of 
the regression coefficients do not exist (i.e., are infinite).
Removing certain covariates from the regression model 
may appear to be an easy remedy for the problem of separation, 
but this ad-hoc strategy has been shown to 
often result in the removal of covariates with 
strong relationships with the response \citep{Zorn_2005}.  

In the frequentist literature, various solutions 
based on penalized or modified likelihoods have been proposed 
to obtain finite parameter estimates
\citep{Firth_1993, Heinze_Schemper_2002, Heinze_2006, Rousseeuw_Christmann_2003}.
The problem has also been noted when fitting 
Bayesian logistic regression models \citep{Clogg_etal_1991}, 
where posterior inferences would be similarly affected by the problem of
separation if using improper priors, with the possibility of 
improper posterior distributions \citep{Speckman_etal_2009}.

\citet{Gelman_etal_2008} recommended using independent Cauchy prior distributions 
as a default weakly informative choice for the regression coefficients in a logistic regression model,
because these heavy tailed priors avoid over-shrinking large
coefficients, but provide shrinkage (unlike improper uniform priors) that enables inferences even 
in the presence of complete separation. 
\citet{Gelman_etal_2008} developed an approximate EM algorithm to obtain the posterior 
mode of regression coefficients with Cauchy priors. While inferences based on
the posterior mode are convenient, often other 
summaries of the posterior distribution are also of interest.
For example, posterior means under Cauchy priors estimated
via Monte Carlo and other approximations have been reported in 
\citet{Bardenet_etal_2014, Chopin_Ridgway_2015}.
It is well-known that the mean does not exist for the Cauchy distribution, 
so clearly the prior means of the regression coefficients do not exist. 
In the presence of separation, where the maximum likelihood estimates 
are not finite, it is not clear whether the posterior means will exist. 
To the best of our knowledge, there has been no investigation considering 
the existence of the posterior mean under Cauchy priors and 
our research is filling this gap. We find a necessary and sufficient
condition where the use of independent Cauchy priors will result in 
finite posterior means here. In doing so we provide further theoretical
underpinning of the approach recommended by \cite{Gelman_etal_2008}, 
and additionally provide further insights on their suggestion of centering 
the covariates before fitting the regression model, 
which can have an impact on the existence of posterior means. 

When the conditions for existence of the posterior mean are satisfied, 
we also empirically compare different prior choices 
(including the Cauchy prior) through various simulated and real data examples.
In general, posterior computation for logistic regression is 
known to be more challenging than probit regression. 
Several MCMC algorithms for logistic regression have been proposed
\citep{OBrien_Dunson_2004, Holmes_Held_2006, Gramacy_Polson_2012},
while the most recent P{\'o}lya-Gamma data augmentation scheme 
of \citet{Polson_etal_2013} emerged superior to the other methods. 
Thus we extend this P{\'o}lya-Gamma Gibbs sampler for normal priors  
to accommodate independent Student-$t$ priors and provide 
an R package to implement the corresponding Gibbs sampler.

The remainder of this article is organized as follows. In Section 2 
we derive the theoretical results:
a necessary and sufficient condition for the existence of
posterior means for coefficients under independent Cauchy priors in a
logistic regression model in the presence of separation, and extend
our investigation to binary regression models with other link functions such as
the probit link, and multivariate Cauchy priors. 
In Section 3 we develop a Gibbs sampler 
for the logistic regression model under independent 
Student-$t$ prior distributions 
(of which the Cauchy distribution is a special case) and briefly
describe the NUTS algorithm of \citet{Hoff:Gelm:2014} which forms the
basis of the software Stan. 
In Section 4 we
illustrate via simulated data that Cauchy priors may lead to 
coefficients of extremely 
large magnitude under separation, 
accompanied by slow mixing Gibbs samplers, 
compared to lighter tailed priors such as 
Student-$t$ priors with degrees of freedom $7$ ($t_7$) or normal priors.
In Section 5 we compare Cauchy, $t_7$, and normal priors based on 
two real datasets, the SPECT data with quasicomplete separation
 and the Pima Indian Diabetes data without separation.
Overall, Cauchy priors exhibit slow mixing under the Gibbs
 sampler compared to the other two priors. Although mixing can
 be improved by the NUTS algorithm in Stan, normal priors seem to be the most preferable 
in terms of producing more reasonable
scales for posterior samples of the regression
coefficients accompanied by competitive predictive performance, under separation.
In Section 6 we conclude with a discussion and our recommendations.

 \section{Existence of Posterior Means Under Cauchy Priors}
 In this section, we begin with a
review of the concepts of complete and quasicomplete separation proposed by 
\citet{Albert_Anderson_1984}. Then based on a new concept of solitary separators, we 
introduce the main theoretical result of this paper, a necessary and sufficient condition 
for the existence of posterior means of regression coefficients under independent Cauchy
priors in the case of separation. 
Finally, we extend our investigation to binary regression models with
other link functions, and Cauchy priors with different scale parameter structures.

Let $\mathbf{y} = (y_1, y_2, \ldots,y_n)^T$ denote a vector of 
independent Bernoulli response variables with success probabilities $\pi_{1}, \pi_2, \ldots, \pi_{n}$. 
For each of the observations, $i = 1, 2, \ldots, n$,
let $\mathbf{x}_i= (x_{i1}, x_{i2}, \ldots,x_{ip})^T$ denote a vector of $p$ covariates,
whose first component is assumed to accommodate the intercept, 
i.e., $x_{i,1}=1$.
Let $\mathbf{X}$ denote the $n \times p$ design matrix with $\mathbf{x}_i^T$ 
as its $i$th row. We assume that 
the column rank of $\mathbf{X}$ is greater than 1.
In this paper, we mainly focus on the logistic regression model,
which is expressed as:
\begin{equation}
\log\left(\frac{\pi_{i}}{1-\pi_i}\right) = \mathbf{x}_i^T\boldsymbol\beta, \quad i = 1,
2, \ldots, n,
\label{eqn:logreg}
\end{equation}
where $\boldsymbol\beta = (\beta_1, \beta_2, \ldots, \beta_p)^T$ is the vector of
regression coefficients.

\subsection{A Brief Review of Separation}

We denote two disjoint subsets of sample points based on their
response values: $ A_0 = \{i: y_i = 0\}$ and $A_1 = \{i: y_i = 1\}$.
According to the definition of \citet{Albert_Anderson_1984}, 
complete separation occurs in the sample if
there exists a 
vector $\boldsymbol\alpha = (\alpha_1, \alpha_2, \ldots,\alpha_p)^T$,
such that for all $i = 1, 2, \ldots, n$, 
\begin{equation}
\mathbf{x}_i^T \boldsymbol\alpha > 0 \ \text{ if } i \in A_1, \quad 
\mathbf{x}_i^T \boldsymbol\alpha < 0 \ \text{ if } i \in A_0. \label{eqn:compsep}
\end{equation}
Consider a simple example in which we wish to predict whether subjects in a study have a
certain kind of infection based on model \eqref{eqn:logreg}. Let $ y_i = 1$ if the $i$th subject is infected and 0 otherwise. 
The model includes an intercept ($x_{i1}=1$) and the other covariates are age ($x_{i2}$), gender ($x_{i3}$), and previous records of being infected 
($x_{i4}$). Suppose in the sample, all infected subjects are older than 25 ($x_{i2} > 25$), and all subjects who are
not infected are younger than 25 ($x_{i2} < 25$). This is an example of complete separation because \eqref{eqn:compsep} is satisfied
for $\boldsymbol\alpha = (-25,1,0,0)^T$.

If the sample points cannot be completely separated, 
\citet{Albert_Anderson_1984} introduced another notion of separation called 
quasicomplete separation. There is quasicomplete separation in the sample
if there exists a non-null vector $\boldsymbol\alpha = (\alpha_1, \alpha_2, \ldots, \alpha_p)^T$,
such that for all $i = 1, 2, \ldots, n$, 
\begin{equation}
\mathbf{x}_i^T \boldsymbol\alpha \geq 0 \ \text{ if } i \in A_1, \quad
\mathbf{x}_i^T \boldsymbol\alpha \leq 0 \ \text{ if } i \in
A_0,   \label{eqn:qcompsep}
\end{equation}
and equality holds for at least one $i$.
Consider the set up of the previous example where the goal is to predict whether a person is infected or not. 
Suppose we have the same model but there is a slight modification in the dataset: all infected subjects are at least 25 years old ($x_{i2} \geq 25$), all uninfected 
subjects are no more than 25 years old ($x_{i2} \leq 25$), and there are two subjects aged exactly 25, of whom one is infected but not the other. 
This is an example of quasicomplete separation because \eqref{eqn:compsep} is satisfied
for $\boldsymbol\alpha = (-25,1,0,0)^T$ and the equality holds for two observations with age exactly 25.

Let $\mathcal{C}$ and $\mathcal{Q}$ denote the set of all vectors $\bm{\alpha}$ that satisfy \eqref{eqn:compsep} and \eqref{eqn:qcompsep}, respectively.
For any  $\bm{\alpha} \in \mathcal{C}$, all sample points must satisfy \eqref{eqn:compsep},
so $\bm{\alpha}$ cannot lead to quasicomplete separation which requires at least one equality in  \eqref{eqn:qcompsep}. 
This implies that $\mathcal{C}$ and $\mathcal{Q}$ are disjoint sets,
while both can be non-empty for a certain dataset. Note that
\citet{Albert_Anderson_1984} define quasicomplete separation only when 
the sample points cannot be separated using complete separation. Thus 
according to their definition, only one of $\mathcal{C}$ and $\mathcal{Q}$ can be
non-empty for a certain dataset. However, 
in our slightly modified definition of quasicomplete separation, 
the absence of complete separation is not required. This permits
both $\mathcal{C}$ and $\mathcal{Q}$ to be non-empty for a dataset.
In the remainder of the paper, for simplicity we use the term ``separation''
to refer to either complete or quasicomplete separation, so that
$\mathcal{C} \cup \mathcal{Q}$ is non-empty.

\subsection{Existence of Posterior Means Under Independent Cauchy Priors}

When Markov chain Monte Carlo (MCMC) is applied to sample from
the posterior distribution, the posterior mean is a commonly used
summary statistic. 
We aim to study whether the marginal posterior mean
$E(\beta_j \mid \mathbf{y})$ exists under the independent Cauchy priors
suggested by \citet{Gelman_etal_2008}. 
Let $C(\mu,\sigma)$ denote a Cauchy distribution with location
parameter $\mu$ and scale parameter $\sigma$. 
The default prior suggested by \citet{Gelman_etal_2008} corresponds to
$\beta_j \stackrel{\text{ind}}{\sim} C(0,\sigma_j)$, for $j=1, 2, \dots, p$.

For a design matrix with full column rank, \citet{Albert_Anderson_1984}
showed that a finite maximum likelihood estimate of $\boldsymbol{\beta}$
does not exist when there is separation in the data.
However, even in the case of separation and/or a rank deficient design
 matrix, the posterior means for some or all $\beta_j$'s may
 exist because they 
incorporate the information from the prior distribution. 
Following Definition 2.2.1 of \citet[pp.\ 55]{Casella_Berger_1990},
we say $E(\beta_j \mid \mathbf{y})$ exists if 
$E(|\beta_j| \mid \mathbf{y}) < \infty$, and in this case,
$E(\beta_j \mid \mathbf{y})$ is given by
\begin{equation}\label{eqn:pm_betaj}
E(\beta_j \mid \mathbf{y})
 = \int_0^{\infty} \beta_j ~p(\beta_j \mid \mathbf{y})~d\beta_j  
 + \int_{-\infty}^0 \beta_j ~p(\beta_j \mid \mathbf{y})~d\beta_j.
\end{equation}
Note that alternative definitions may require
only one of the integrals in \eqref{eqn:pm_betaj} to be
finite for the mean to exist, e.g., \citet[pp.\ 455]{Bickel_Doksum_2001}. 
However, according to the definition used in this paper, both integrals in
\eqref{eqn:pm_betaj} have to be finite for the posterior mean to
exist.
Our main result shows that for each $j = 1, 2, \ldots, p$,
the existence of $E(\beta_j \mid \mathbf{y})$ depends on whether the predictor $\mathbf{X}_j$ 
is a solitary separator or not, which is defined as follows:

\begin{definition}
The predictor $\mathbf{X}_j$ 
is a solitary separator, if there exists an $\boldsymbol\alpha \in (\mathcal{C} \cup \mathcal{Q})$ such that
\begin{equation}\label{eqn:solitary}
\alpha_j \neq 0, \quad \alpha_r = 0 \text{ for all } r \neq j.
\end{equation}
\end{definition}

This definition implies that for a solitary separator $\mathbf{X}_j$, 
if $\alpha_j > 0$, then $x_{i,j} \geq 0$ for all $i\in A_1$, and $x_{i,j} \leq 0$ for all $i \in A_0$;
if $\alpha_j < 0$, then $x_{i,j} \leq 0$ for all $i\in A_1$, and $x_{i,j} \geq 0$ for all $i \in A_0$.
Therefore, the hyperplane $\{\mathbf{x}\in \mathbb{R}^p: x_{j} = 0\}$ in the predictor space 
separates the data into two groups $A_1$ and $A_0$ 
(except for the points located on the hyperplane).
The following theorem provides a necessary and sufficient
condition for the existence of marginal posterior means of
  regression coefficients in a logistic regression model. 
\begin{theorem}\label{theorem:existence}
In a logistic regression model, suppose the regression coefficients 
(including the intercept) $\beta_j \stackrel{ind}{\sim} C(0,\sigma_j)$
with $\sigma_j>0$ for $j=1, 2, \dots, p$, so that
\begin{equation}\label{eqn:prior}
p(\bm{\beta}) = \prod_{j=1}^{p} p(\beta_j)
= \prod_{j=1}^{p}\frac{1}{\pi \sigma_j (1 + \beta_j^2 / \sigma_j^2)}. 
\end{equation}
Then for each $j = 1, 2, \ldots, p$, the posterior mean $E(\beta_j \mid \mathbf{y})$ exists if and only if 
$\mathbf{X}_j$ is not a solitary separator.
\end{theorem}

A proof of Theorem \ref{theorem:existence} is available in Appendices
\ref{proof_th1_necessary} and \ref{proof_th1_sufficient}.

\begin{remark}
Theorem \ref{theorem:existence} implies that under independent Cauchy
priors in logistic regression,
the posterior means of all coefficients exist if there is no separation, 
or if there is separation with no solitary separators.
\end{remark}
\begin{remark} 
\citet{Gelman_etal_2008} suggested centering all predictors (except interaction terms) 
in the pre-processing step. A consequence of Theorem \ref{theorem:existence} is
that centering may have a crucial role in the existence of 
the posterior mean $E(\beta_j \mid \mathbf{y})$.
\end{remark}

We expand on the second remark with a  toy example where a predictor is a solitary
separator before centering but not after centering.
Consider a dataset with $n=100$, 
$\mathbf{y}=(\underbrace{0, \dots 0}_{25}, \underbrace{1, \dots,
  1}_{75})^T$ and a binary predictor 
$\mathbf{X}_j = (\underbrace{0, \dots 0}_{50}, \underbrace{1, \dots, 1}_{50})^T$. 
Here $\mathbf{X}_j$ is a solitary separator which leads to
quasicomplete separation before centering. However, the centered predictor 
$\mathbf{X}_j = (\underbrace{-0.5, \dots -0.5}_{50}, \underbrace{0.5, \dots, 0.5}_{50})^T$
is no longer a solitary separator because 
after centering the hyperplane $\{\mathbf{x}: x_{j} = -0.5\}$ separates the data but 
$\{\mathbf{x}: x_{j} = 0\}$ does not. 
Consequently, the posterior mean $E(\beta_j \mid \mathbf{y})$ does not
exist before centering but it exists after centering.

\subsection{Extensions of the Theoretical Result}

So far we have mainly focused on the logistic regression model, which is
one of the most widely used binary regression models because of
the interpretability of its regression coefficients in terms of
odds ratios. We now generalize Theorem \ref{theorem:existence} 
to binary regression models with link functions other than the logit.
Following the definition in \citet[pp.\ 27]{McCullagh_Nelder_1989},
we assume that for $i = 1, 2, \ldots, n$,
the linear predictor $\mathbf{x}_i^T\boldsymbol\beta$ 
and the success probability $\pi_{i}$ are connected by a
monotonic and differentiable link function $g(\cdot)$
such that $g(\pi_{i}) = \mathbf{x}_i^T\boldsymbol\beta$.
We further assume that $g(.)$ is a one-to-one function, which means that
  $g(.)$ is strictly monotonic. This
is satisfied by many commonly used link functions including the probit.
Without loss of generality, we assume that $g(\cdot)$ is a
 strictly increasing function.

\begin{theorem}\label{theorem:general}
In a binary regression model with link function $g(.)$ described above, suppose the regression 
coefficients have  independent Cauchy priors in \eqref{eqn:prior}.
Then for each $j =1, 2, \ldots, p$, 
\begin{itemize}
\item[(1)] a necessary condition for the existence of the posterior mean
$E(\beta_j \mid \mathbf{y})$ is that $\mathbf{X}_j$ is not a solitary separator; 
\item[(2)] a sufficient condition for the existence of
  $E(\beta_j \mid \mathbf{y})$ consists of the following:
\begin{itemize}
\item[(i)] $\mathbf{X}_j$ is not a solitary separator, and
\item[(ii)] $\forall \epsilon>0$, 
\begin{equation}\label{eq:general_condition}
 \int_{0}^{\infty} \beta_j p(\beta_j) g^{-1}(- \epsilon \beta_j) d \beta_j < \infty,
\quad
\int_{0}^{\infty} \beta_j p(\beta_j) \left[1 - g^{-1}(\epsilon \beta_j)\right] d \beta_j< \infty.
\end{equation}
\end{itemize}
\end{itemize}
\end{theorem}
Note that \eqref{eq:general_condition} in the sufficient
condition of Theorem \ref{theorem:general} imposes constraints 
on the link function $g(.)$, and hence the likelihood function. 
A proof of this theorem is given in Appendix \ref{proof_th2}.
Moreover, it is shown that condition \eqref{eq:general_condition}
holds for the probit link function.

In certain applications, to incorporate available prior information, 
it may be desirable to use Cauchy priors with nonzero location parameters. 
The following corollary states that for both logistic 
and probit regression, the condition for existence of posterior means derived 
in Theorems \ref{theorem:existence} and \ref{theorem:general}
continues to hold under independent Cauchy priors with nonzero location parameters.

\begin{corollary}\label{corollary:nonzero_prior_mean}
In logistic and probit regression models, suppose the  regression coefficients 
$\beta_j \stackrel{\text{ind}}{\sim} C(\mu_j,\sigma_j)$, for $j=1, 2, \dots, p$.
Then a necessary and sufficient condition for the existence of the posterior mean 
$E(\beta_j \mid \mathbf{y})$ is that $\mathbf{X}_j$ is not a solitary separator, for $j=1, 2, \dots, p$.
\end{corollary}

A proof of Corollary \ref{corollary:nonzero_prior_mean} is available in Appendix 
\ref{proof_corollary:nonzero_prior_mean1}.

In some applications it could be more natural to allow the regression coefficients to be dependent, 
{\it a priori}. Thus
in addition to independent Cauchy priors, we also study
the existence of posterior means under a multivariate Cauchy prior, with the following density function:
\begin{equation}\label{eq:multi_Cauchy_prior}
p(\boldsymbol\beta) = \frac{\Gamma\left( \frac{1+p}{2} \right)}
	{\Gamma\left( \frac{1}{2} \right) \pi^{\frac{p}{2}} |\boldsymbol\Sigma|^{\frac{1}{2}}
	\left[ 1 + (\boldsymbol\beta - \boldsymbol\mu)^T \boldsymbol\Sigma^{-1}
	(\boldsymbol\beta - \boldsymbol\mu) \right]^{\frac{1 + p}{2}}},
\end{equation}
where $\boldsymbol\beta \in \mathbb{R}^p$, $\boldsymbol\mu$ is a $p \times 1$ location parameter
and $\boldsymbol\Sigma$ is a $p \times p$ positive-definite scale matrix.
A special case of the multivariate Cauchy prior is the Zellner-Siow prior \citep{Zellner_Siow_1980}.
It can be viewed as a scale mixture of $g$-priors, where conditional on $g$,
$\boldsymbol\beta$ has a multivariate normal prior with
a covariance matrix proportional to $g(\mathbf{X}^T\mathbf{X})^{-1}$,
and the hyperparameter $g$ has an inverse gamma prior, $\text{IG}(1/2, n/2)$.
Based on generalizations of the $g$-prior to binary regression models
\citep{Fouskakis_etal_2009, Bove_Held_2011, Hanson_etal_2014},
the Zeller-Siow prior, which has a density
\eqref{eq:multi_Cauchy_prior} with $\boldsymbol\Sigma \propto n(\mathbf{X}^T\mathbf{X})^{-1}$,
can be a desirable objective prior as it preserves the covariance structure 
of the data and is free of tuning parameters.

\begin{theorem}\label{theorem:multivariate_Cauchy}
In logistic and probit regression models, suppose the vector of regression coefficients 
$\boldsymbol\beta$ has a multivariate Cauchy prior as in \eqref{eq:multi_Cauchy_prior}.
If there is no separation, then all posterior means $E(\beta_j \mid \mathbf{y})$ exist,
for $j = 1, 2, \ldots, p$. If there is complete separation, then none of the posterior means 
$E(\beta_j \mid \mathbf{y})$ exist, for  $j = 1, 2, \ldots, p$.
\end{theorem}

\begin{figure}[h!]
\begin{center}
\includegraphics[height=2.15in,width=4in,angle=0]{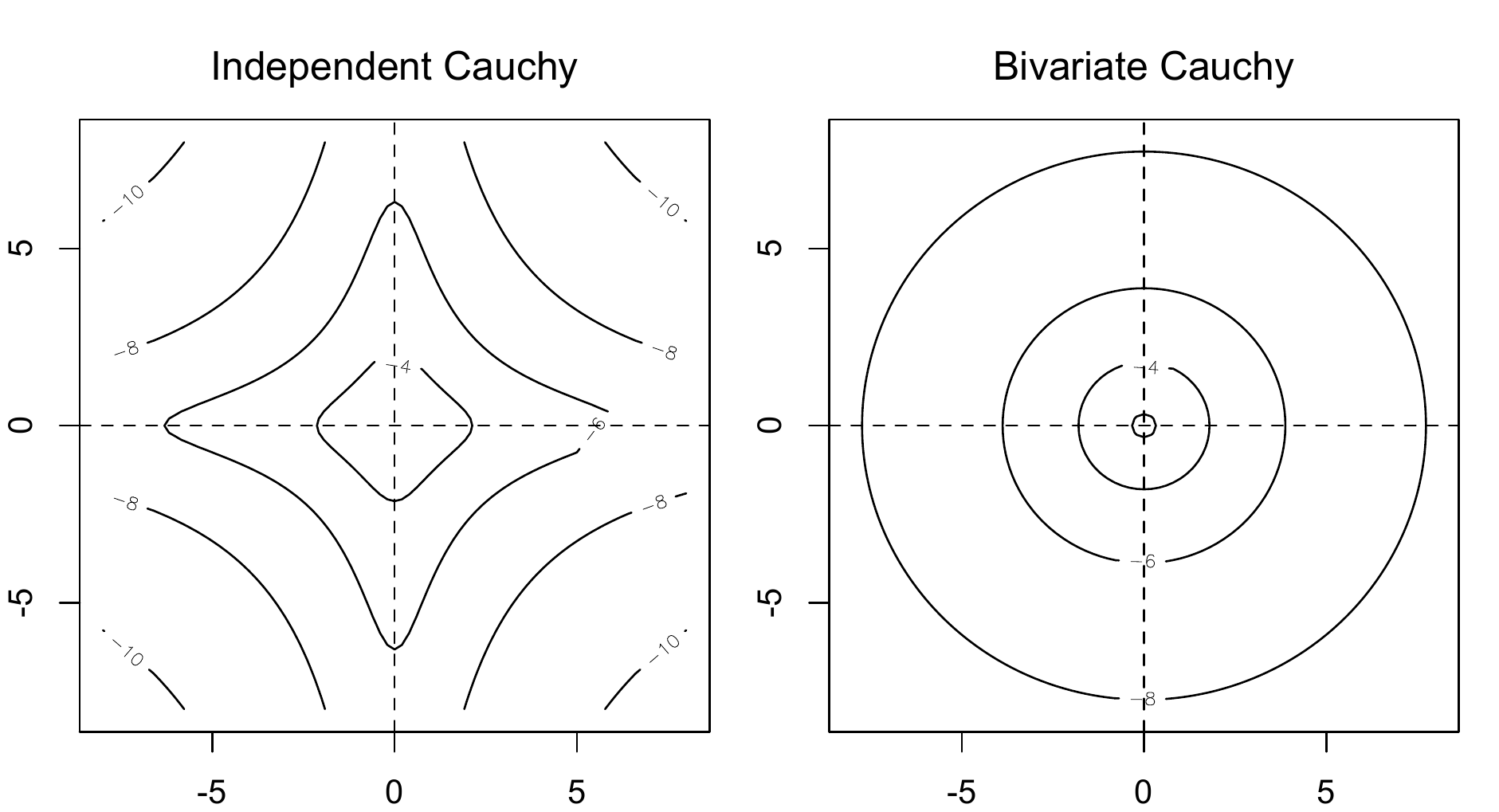}
\end{center}
\caption{Contour plots of log-density functions of 
independent Cauchy distributions with both scale parameters being $1$ (left)
and a bivariate Cauchy distribution with scale matrix $\mathbf{I}_2$ (right).
These plots suggest that independent Cauchy priors place more
probability mass along axes than a multivariate Cauchy prior, and thus
impose stronger shrinkage.
Hence, if complete separation occurs, 
$E(\beta_j \mid \mathbf{Y})$ may exist under independent Cauchy
priors for some or all $j = 1,2, \ldots, p$ (Theorem \ref{theorem:existence}),
but does not exist under a 
multivariate Cauchy prior (Theorem \ref{theorem:multivariate_Cauchy}).}
\label{fig:Cauchy_contours}
\end{figure}

A proof of Theorem \ref{theorem:multivariate_Cauchy} is available in Appendices
\ref{proof_theorem:multivariate_Cauchy_no_separation} and
\ref{proof_theorem:multivariate_Cauchy_complete_separation}.
The study of existence of posterior means under multivariate Cauchy priors
in the presence of quasicomplete separation has proved to be more challenging.
We hope to study this problem in future work.
Note that although under \eqref{eq:multi_Cauchy_prior},
the induced marginal prior of $\beta_j$ is a univariate Cauchy distribution for
each $j = 1, 2, \ldots, p$, the multivariate Cauchy prior is different from independent Cauchy priors,
even with a diagonal scale matrix $\boldsymbol\Sigma = \text{diag}(\sigma_1^2, \sigma_2^2,\ldots, \sigma_p^2)$.
In fact, as a rotation invariant distribution, the multivariate Cauchy prior places less
probability mass along axes than the independent Cauchy priors (see Figure \ref{fig:Cauchy_contours}).
Therefore, it is not surprising that solitary separators no longer play an important 
role for existence of posterior means under multivariate Cauchy priors, as evident from Theorem \ref{theorem:multivariate_Cauchy}.

So far we have considered Cauchy priors, which are $t$ distributions with 1 degree
of freedom. We close this section with a remark on lighter tailed $t$ priors
(with degrees of freedom greater than 1) and normal priors, for which the prior means exist.
\begin{remark}
In a binary regression model, suppose that the regression coefficients 
have independent Student-t priors with degrees of freedom greater
than one, or independent normal priors.
Then it is straightforward to show that the posterior means of the
coefficients exist
because the likelihood is bounded above by one and the prior means exist.
The same result holds under multivariate $t$ priors with degrees
of freedom greater than one, and multivariate normal priors.
\end{remark}

\section{MCMC Sampling for Logistic Regression}\label{sec:mcmc}
In this section we discuss two algorithms for sampling from the posterior distribution
for logistic regression coefficients under independent Student-$t$ priors. We first develop a Gibbs sampler and then 
briefly describe the No-U-Turn Sampler (NUTS) implemented in the freely available software Stan \citep{Carp:Gelm:Hoff:Lee:Good:Beta:Brub:Guo:Li:Ridd:2016}.

\subsection{P{\'o}lya-Gamma Data Augmentation Gibbs Sampler}\label{sec:gibbs}
\citet{Polson_etal_2013} showed that the likelihood for 
logistic regression can be written as a mixture of normals with respect to 
a P{\'o}lya-Gamma (PG) distribution.
Based on this result, they developed an efficient Gibbs sampler for 
logistic regression with a multivariate normal prior on $\boldsymbol\beta$.
\citet{Choi_Hobert_2013} showed that their Gibbs sampler
is uniformly ergodic. This guarantees the existence of 
central limit theorems for Monte Carlo averages of 
functions of $\boldsymbol\beta$ which are square integrable 
with respect to the posterior distribution $p(\boldsymbol\beta \mid \mathbf{y})$.
\citet{Choi_Hobert_2013} developed a latent data model 
which also led to the Gibbs sampler of \citet{Polson_etal_2013}. 
We adopt their latent data formulation to develop a Gibbs sampler
for logistic regression with independent Student-$t$ priors on 
$\boldsymbol\beta$.

Let $U = (2 / \pi^2)\sum_{l=1}^{\infty}W_{l}/{(2l-1)}^2$, 
where $W_{1},W_{2}, \dots$ is a sequence of i.i.d.~Exponential random variables
with rate parameters equal to 1. The density of $U$ is given by
\begin{equation}
h(u) = \sum_{l=0}^{\infty}{(-1)}^l \frac{(2l+1)}{\sqrt{2 \pi u^3}}e^{-\frac{(2l+1)^2}{8u}}, \quad 0< u < \infty. 
\end{equation}
Then for $k \geq 0$, the P{\'o}lya-Gamma (PG) distribution is constructed by exponential tilting of $h(u)$ as follows:
\begin{equation}
p(u; k) = \cosh\left(\frac{k}{2}\right) e^{-\frac{k^2u}{2}}h(u), \quad 0< u <\infty.
\end{equation}
A random variable with density $p(u; k)$ has a PG($1,k$) distribution.

Let $ t_{v}(0, \sigma_j)$ denote the Student-$t$ distribution with $v$ degrees of freedom, location parameter 0, and scale parameter $\sigma_j$.
Since Student-$t$ distributions can be expressed as 
inverse-gamma (IG) scale mixtures of normal distributions, 
for $j = 1, 2, \ldots, p$, we have:
\begin{equation*}
\beta_j \sim t_{v}(0, \sigma_j) \Longleftrightarrow
\begin{cases}
\beta_j \mid \gamma_j \sim \text{N}(0, \gamma_j), \\
\gamma_j \sim \text{IG}\left(\frac{v}{2}, \frac{v\sigma_j^2}{2}  \right).
\end{cases}
\end{equation*}
Conditional on $\boldsymbol\beta$ and 
$\boldsymbol\Gamma = \text{diag}(\gamma_1, \gamma_2, \ldots, \gamma_p)$, let
$(y_1, z_1), (y_2, z_2), \ldots, (y_n, z_n)$ be $n$ independent random vectors 
such that $y_i$ has a Bernoulli distribution with success probability 
$\exp(\mathbf{x}_i^T\boldsymbol\beta)/
(1+\exp(\mathbf{x}_i^T\boldsymbol\beta))$,
$z_i \sim PG(1,|\mathbf{x}_i^T\boldsymbol\beta|)$, 
and $y_i$ and $z_i$ are independent, for $i=1, 2, \ldots, n$.
Let $\boldsymbol{Z}_{D}= \text{diag}(z_1, z_2, \ldots,  z_n)$,
then the augmented posterior density is 
$p(\boldsymbol\beta,\boldsymbol\Gamma,\boldsymbol{Z}_{D} \mid \mathbf{y})$.
We develop a Gibbs sampler with target distribution $p(\boldsymbol\beta,
\boldsymbol\Gamma,\boldsymbol{Z}_{D} \mid \mathbf{y})$,
which cycles through the following sequence of distributions iteratively:
\begin{enumerate}
\item $\boldsymbol\beta \mid \boldsymbol\Gamma, \boldsymbol{Z}_{D}, \mathbf{y}
	\sim \text{N}\left( (\mathbf{X}^T\boldsymbol{Z}_{D}\mathbf{X} + \boldsymbol\Gamma^{-1})^{-1}
	\mathbf{X}^T\mathbf{\tilde{y}},
	(\mathbf{X}^T\boldsymbol{Z}_{D}\mathbf{X} + \boldsymbol\Gamma^{-1})^{-1} \right)$, where 
 $\tilde{y}_i = y_i -1/2$ and $\mathbf{\tilde{y}} = (\tilde{y}_1, \tilde{y}_2, \ldots, \tilde{y}_n)^T$,
\item $\gamma_j \mid \boldsymbol\beta, \boldsymbol{Z}_{D}, \mathbf{y}
	 \stackrel{\text{ind}}{\sim} \text{IG}\left( \frac{v+1}{2}, \frac{\beta_j^2 + v\sigma_j^2}{2} \right)$, for $j=1, 2, \ldots, p$,
\item $z_i \mid  \boldsymbol\Gamma,\boldsymbol\beta,\mathbf{y}
	 \stackrel{\text{ind}}{\sim} \text{PG}(1, |\mathbf{x}_i^T\boldsymbol\beta|)$, for $i=1, 2, \ldots, n$.	
\end{enumerate}

Steps 1 and 3 follow immediately from \citet{Choi_Hobert_2013, 
Polson_etal_2013} and step 2 follows from straightforward algebra.
In the next section, for comparison of posterior distributions 
under Student-$t$ priors with different degrees of freedom, 
we implement the above Gibbs sampler, and
for normal priors we apply the Gibbs sampler of \citet{Polson_etal_2013}. 
Both Gibbs samplers can be implemented using the R package {\tt tglm},
available in the supplement.   

\subsection{Stan}\label{sec:nuts}
Our empirical results in the next section suggest that the Gibbs sampler exhibits
extremely slow mixing for posterior simulation under Cauchy
priors for data with separation.
Thus we consider alternative MCMC sampling algorithms in the hope of improving mixing. 
A random walk Metropolis algorithm shows some improvement over the Gibbs sampler in the $p=2$ case. 
However, it is not efficient for exploring higher dimensional spaces.
 Thus we have been motivated to use the software Stan 
\citep{Carp:Gelm:Hoff:Lee:Good:Beta:Brub:Guo:Li:Ridd:2016}, which 
implements the No-U-Turn Sampler (NUTS) of \citet{Hoff:Gelm:2014}, a 
tuning free extension of the Hamiltonian Monte Carlo (HMC) algorithm \citep{Neal:2011}.

It has been demonstrated that for continuous parameter spaces, HMC can improve over poorly mixing Gibbs samplers and random walk Metropolis algorithms.
HMC is a Metropolis algorithm  
that generates proposals  
based on Hamiltonian dynamics, a concept borrowed from Physics.
In HMC, the parameter of interest is referred to as the ``position''
variable, representing a particle's position in a $p$-dimensional
space. A $p$-dimensional auxiliary parameter, the ``momentum'' variable,
is introduced to represent the particle's momentum. 
In each iteration, the momentum variable is 
generated from a Gaussian distribution,  
and then a proposal of the position momentum pair is generated 
(approximately) along the trajectory of the Hamiltonian dynamics
defined by the joint distribution of the position and momentum.  
Hamiltonian dynamics changing
over time can be approximated by discretizing time via the
``leapfrog'' method. In practice, 
a proposal is generated by
applying the leapfrog algorithm $L$ times, with stepsize $\epsilon$, to the the current state.
The proposed state is accepted or rejected according to a Metropolis acceptance probability.
Section 5.3.3 of the review paper by 
\citet{Neal:2011} illustrates the practical benefits of HMC over random walk Metropolis algorithms. 
The examples in this section 
demonstrate that the momentum variable may change only slowly along certain directions 
during leapfrog steps, permitting the position variable to move consistently in this direction 
for many steps. In this way, proposed states using Hamiltonian dynamics 
can be far away from current states but still 
achieve high acceptance probabilities, making HMC 
more efficient than traditional algorithms such as random walk Metropolis.

In spite of its advantages, HMC has not been very widely used in the Statistics community until recently, 
because its performance can be sensitive to the choice of two tuning
parameters: the leapfrog stepsize $\epsilon$ and the number of leapfrog steps $L$. 
Very small
$\epsilon$ can lead to waste in computational power whereas large
$\epsilon$ can yield large errors due to discretization. 
Regarding the number of leapfrog steps $L$, if it is too small, 
proposed states can be near current states and thus resemble random walk.
On the other hand, if $L$ is too large, the Hamiltonian trajectory can retrace its path so
that the proposal is brought closer to the current value, which
again is a waste of computational power.

The NUTS algorithm tunes these two parameters automatically. 
To select $L$, the main idea is to run the leapfrog steps until the trajectory starts to 
retrace its path. 
More specifically, 
NUTS builds a binary tree based on a recursive doubling
procedure, that is similar in flavor to the doubling procedure used for slice sampling
by \citet{Neal:2003}, with nodes of the tree representing position momentum
pairs visited by the leapfrog steps along the path. 
The doubling procedure is stopped if the trajectory 
starts retracing its path, that is making a ``U-turn'', or if
there is a large simulation error accumulated due to many steps of leapfrog discretization. 
NUTS consists of a carefully constructed
transition kernel that leaves the target joint distribution invariant. 
It also proposes a way for adaptive tuning of
the stepsize $\epsilon$.

We find that by implementing this tuning free NUTS algorithm,
available in the freely available software Stan,  substantially better mixing 
than the Gibbs sampler can be achieved in all of our examples in which posterior
means exist. 
We still include the Gibbs sampler in this article for two main reasons. First, it illustrates that Stan can provide an incredible improvement in mixing 
over the Gibbs sampler in certain cases. Stan requires minimal coding effort, much less than developing a Gibbs sampler, 
which may be useful information for readers who are not yet familiar with Stan. 
Second, Stan currently works for continuous target distributions only, 
but discrete distributions for models and mixed distributions for regression coefficients 
frequently arise in Bayesian variable selection, for regression models with binary or categorical response variables 
\citep{Holmes_Held_2006,Mitr:Duns:2010,Ghos:Clyd:2011,Ghos:Herr:Sieg:2011,Ghos:Reit:2013,Li:Clyd:2015}. Unlike HMC algorithms, Gibbs
samplers can typically be extended via data augmentation to incorporate mixtures of a point mass and a continuous distribution, as priors for the regression coefficients, without much additional 
effort.
\section{Simulated Data}
In this section, we use two simulation examples to empirically demonstrate that 
under independent Cauchy priors, the aforementioned MCMC algorithm 
for logistic regression may suffer from extremely slow mixing in the presence of
separation in the dataset.

For each simulation scenario, we first standardize the predictors
following the recommendation of \citet{Gelman_etal_2008}. 
Binary predictors (with 0/1 denoting the two categories) 
are centered to have mean 0, and other predictors are centered 
and scaled to have mean 0 and standard deviation 0.5. 
Their rationale is that such standardizing 
makes the scale of a continuous predictor comparable to that 
of a symmetric binary predictor, in the sense that they have
the same sample mean and sample standard deviation. 
\citet{Gelman_etal_2008} made a distinction between 
input variables and predictors, and they
suggested standardizing the input variables only. 
For example, temperature and humidity may be input 
variables as well as predictors in a model;
however, their interaction term is a predictor but not an input variable. 
In our examples, except for the constant term for the intercept, 
all other predictors are input variables and standardized appropriately.

We compare the posterior distributions under independent 
i) Cauchy, i.e., Student-$t$ with 1 degree of freedom, 
ii) Student-$t$ with 7 degrees of freedom ($t_{7}$), and 
iii) normal priors for the regression coefficients.
In binary regression models, while the inverse cumulative 
distribution function (CDF) of the logistic distribution 
yields the logit link function, the inverse CDF of the Student-$t$ distribution
yields the robit link function. 
\citet{Liu_2004} showed that the logistic link can be well 
approximated by a robit link with 7 degrees of freedom. So a $t_{7}$ prior
approximately matches the tail heaviness of the logistic likelihood underlying logistic regression.
For Cauchy priors we use the default choice recommended 
by \citet{Gelman_etal_2008}: all location parameters are set to 0 
and scale parameters are set to 10 and 2.5 for the intercept and other coefficients, respectively.
To be consistent we use the same location and scale parameters for the other two priors.
\citet{Gelman_etal_2008} adopted a similar strategy in one of their analyses, 
to study the effect of tail heaviness of the priors.
Among the priors considered here, the normal prior has the lightest tails, 
the Cauchy prior the heaviest, and
the $t_{7}$ prior offers a compromise between the two extremes. 
For each simulated dataset, we
run both the Gibbs sampler developed in Section \ref{sec:gibbs} and Stan, for 1,000,000 iterations
after a burn-in of 100,000 samples, under each of the three priors.

\subsection{Complete Separation with a Solitary Separator}
\begin{figure}[h!]
\begin{center}
\includegraphics[height=3in,width=5.2in,angle=0]{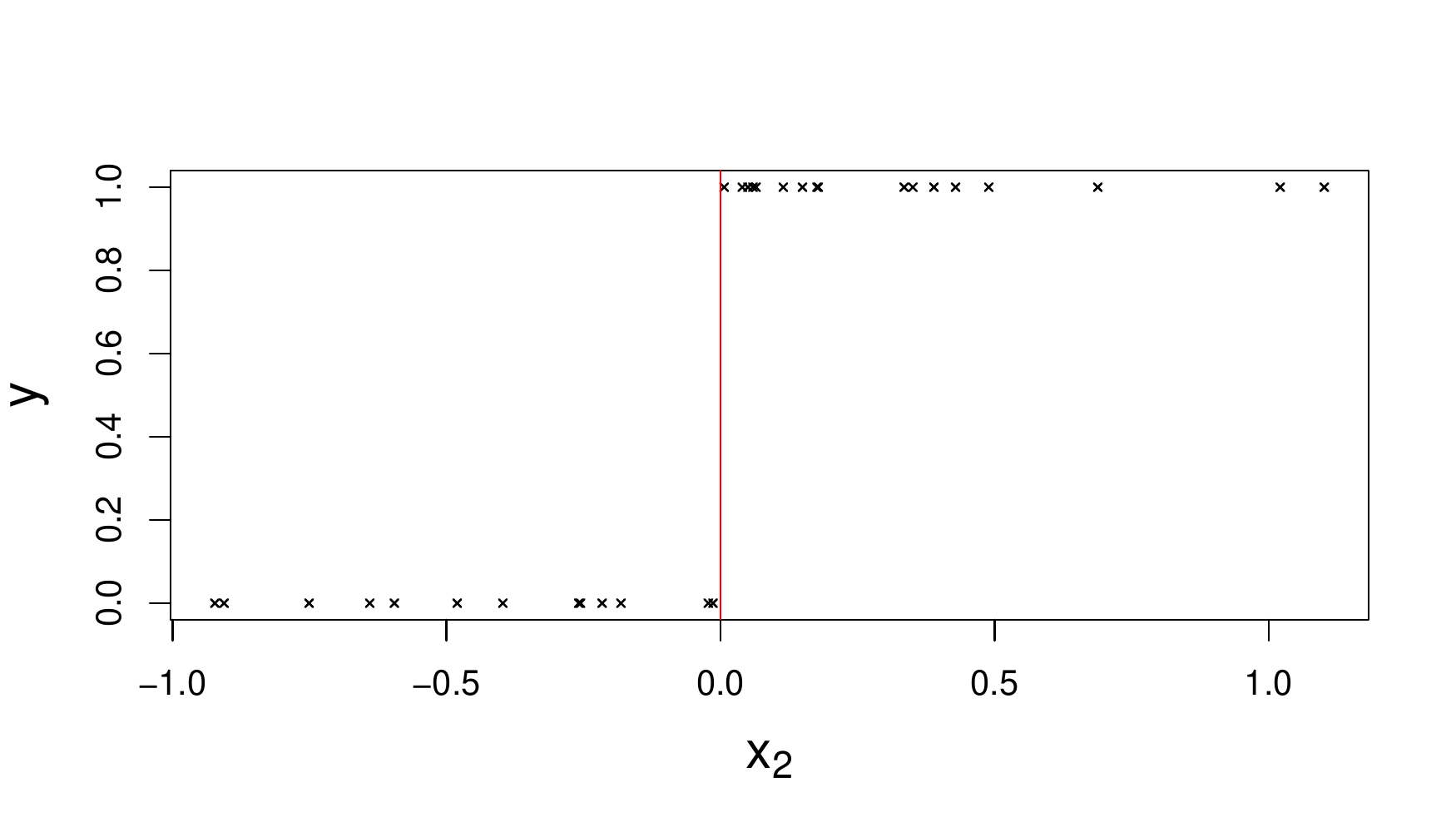}
\end{center}
\caption{Scatter plot of $\mathbf{y}$ versus $\mathbf{X}_2$
in the first simulated dataset, where
$\mathbf{X}_2$ is a solitary separator which completely
 separates the samples (the vertical line at zero separates the points corresponding to
 $y = 1$ and $y = 0$).}
\label{fig:sim1scatter}
\end{figure}

First, we generate a dataset with $p=2$ (including the intercept)
and $n=30$. 
The continuous predictor $\mathbf{X}_2$
is chosen to be a solitary separator (after standardizing), 
which leads to complete separation, whereas the constant term 
$\mathbf{X}_1$ contains all one's and is not a solitary separator.
A plot of $\mathbf{y}$ versus $\mathbf{X}_2$ in 
Figure \ref{fig:sim1scatter} demonstrates this graphically.
So by Theorem \ref{theorem:existence}, under independent Cauchy priors, 
$E(\beta_1 \mid \mathbf{y})$ 
exists but $E(\beta_2 \mid \mathbf{y})$ does not. 

\begin{figure}[h!]
\begin{center}
\includegraphics[height=5in,width=5in,angle=0]{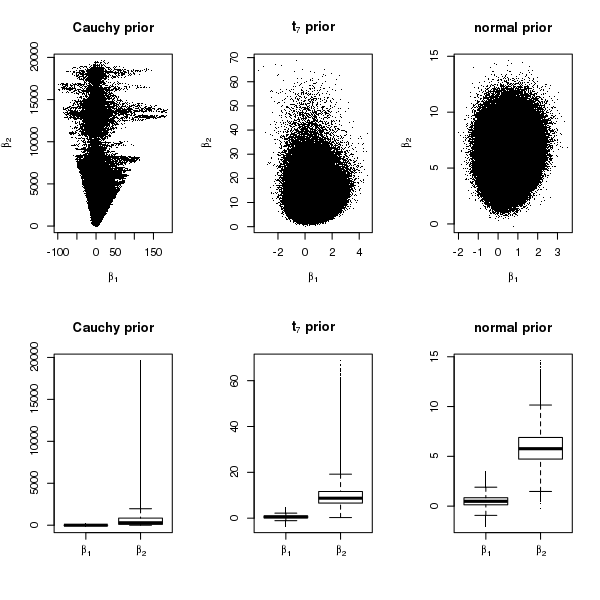}
\end{center}
\caption{Scatter plots (top) and box plots (bottom) of posterior samples
of $\beta_1$ and $\beta_2$ from the Gibbs sampler, under independent Cauchy, $t_7$, and normal priors 
for the first simulated dataset.}
\label{fig:sim1post}
\end{figure}

\begin{figure}[h!]
\begin{center}
\includegraphics[height=5in,width=5in,angle=0]{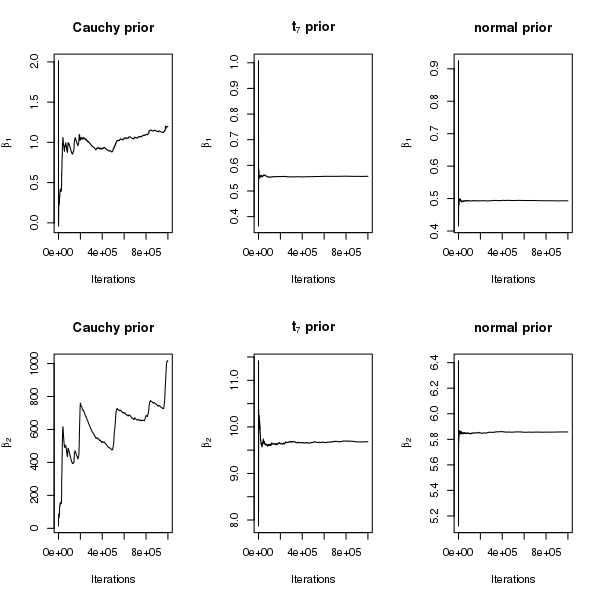}
\end{center}
\caption{Plots of running means of $\beta_1$ (top) and $\beta_2$
(bottom) sampled from the posterior distributions via the Gibbs sampler,
under independent Cauchy, $t_7$, and normal priors for the first simulated dataset. 
Here $E(\beta_1 \mid \mathbf{y})$ exists under independent
Cauchy priors but $E(\beta_2 \mid \mathbf{y})$ does not.}
\label{fig:sim1means}
\end{figure}

The results from the Gibbs sampler are reported in Figures \ref{fig:sim1post} and \ref{fig:sim1means}.
Figure \ref{fig:sim1post} shows the posterior samples
of $\boldsymbol{\beta}$ under the different priors. 
The scale of $\beta_2$, the coefficient corresponding 
to the solitary separator $\mathbf{X}_2$, is extremely large under Cauchy priors, 
less so under $t_7$ priors, and the smallest under normal priors. 
In particular, under Cauchy priors, the posterior distribution of $\beta_2$
seems to have an extremely long right tail. Moreover, although $\mathbf{X}_1$ is not a solitary separator, 
under Cauchy priors, the posterior samples of $\beta_1$ have a 
much larger spread.
Figure \ref{fig:sim1means} shows that
the running means of both $\beta_1$ and $\beta_2$
converge rapidly under normal and $t_7$ priors, 
whereas under Cauchy priors, the running mean of $\beta_1$ 
does not converge after a million iterations and that of $\beta_2$  clearly diverges.
We also ran Stan for this example but do not report the results here, 
because it gave warning messages about divergent transitions for Cauchy priors, after the burn-in period. 
Given that the posterior
mean of $\beta_2$ does not exist in this case, the lack of convergence is not surprising.  

\subsection{Complete Separation Without Solitary Separators}\label{section:sim2}

\begin{figure}[h!]
\begin{center}
\includegraphics[height=3in,width=5.2in,angle=0]{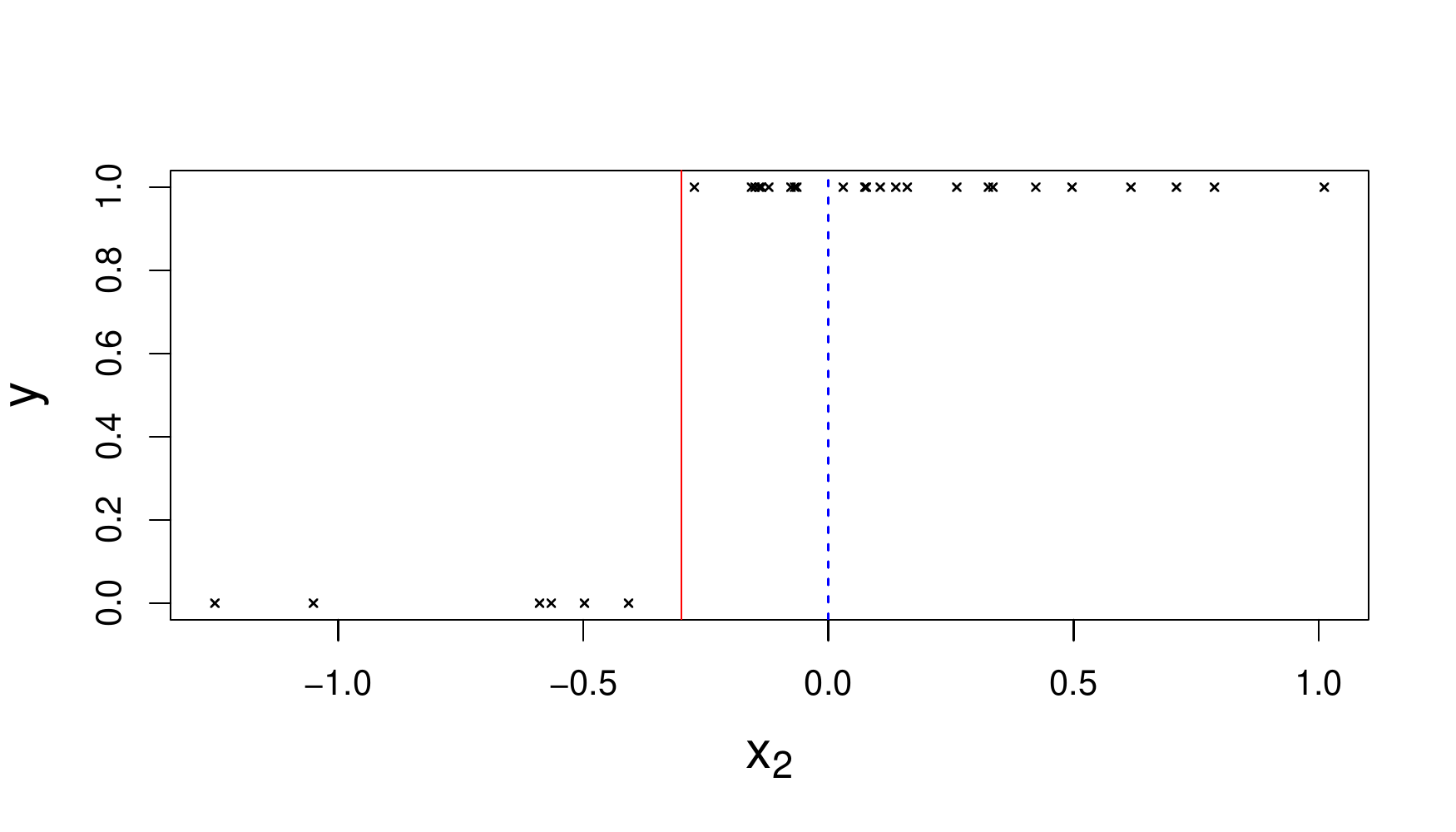}
\end{center}
\caption{Scatter plot of $\mathbf{y}$ versus $\mathbf{X}_2$ 
for the second simulated dataset. The solid vertical line at $-0.3$ demonstrates 
complete separation of the samples.
However, $\mathbf{X}_2$ is not a solitary separator, because
the dashed vertical line at zero does not separate the points 
corresponding to $y = 1$ and $y = 0$. 
The other predictor $\mathbf{X}_1$ is a vector of ones corresponding to 
the intercept, which is not a solitary separator, either. }
\label{fig:sim2scatter}
\end{figure}

\begin{figure}[htb]
\begin{center}
\includegraphics[height=5in,width=5in,angle=0]{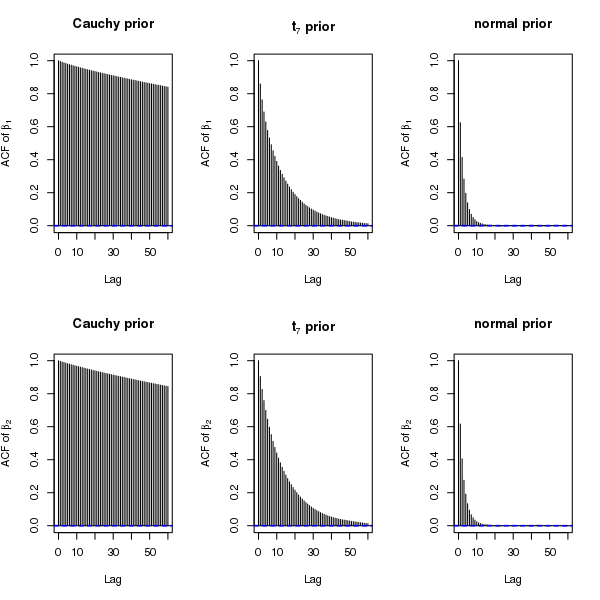}
\end{center}
\caption{Autocorrelation plots of the posterior samples of 
$\beta_1$ (top) and $\beta_2$ (bottom) from the Gibbs sampler, 
under independent Cauchy, $t_7$, and normal priors for the second simulated dataset.}
\label{fig:sim2acf}
\end{figure}

Now we generate a new dataset with $p=2$ and $n=30$ such that 
there is complete separation but there are no solitary separators 
(see Figure \ref{fig:sim2scatter}). 
This guarantees the existence of both $E(\beta_1 \mid \mathbf{y})$ and
$E(\beta_2 \mid \mathbf{y})$ under independent Cauchy priors.
The difference in the existence of $E(\beta_2 \mid \mathbf{y})$
 for the two simulated datasets
is reflected by the posterior samples from the Gibbs sampler: under Cauchy priors,
the samples of $\beta_2$
in Figure 1 in the Appendix 
 are more stabilized than those in 
Figure \ref{fig:sim1post} in the manuscript.
However, when comparing across prior distributions, we find that
the posterior samples of neither $\beta_1$ nor $\beta_2$ are 
as stable as those under $t_7$ and normal priors, which is 
not surprising because among the three priors, Cauchy priors have 
the heaviest tails and thus yield the least shrinkage.
Figure 
2 in the Appendix 
shows that the convergence of the running means 
under Cauchy priors is slow. 
Although we have not verified the existence
of the second or higher order posterior moments under Cauchy priors, 
for exploratory purposes 
we examine sample autocorrelation plots of the draws from the Gibbs sampler. 
Figure \ref{fig:sim2acf} shows that the autocorrelation decays
extremely slowly for Cauchy priors, 
reasonably fast for $t_7$ priors, and rapidly for normal priors. 

Some results from Stan are reported in Figures 3 and 4 in
the Appendix. 
Figure 3 in the Appendix 
shows posterior distributions with nearly identical shapes as those
obtained using Gibbs sampling in Figure 1 in the Appendix, 
 with the only difference being that more extreme values appear under Stan. This is most likely due to faster mixing in Stan. 
As Stan traverses the parameter space more rapidly, values in the
tails appear more quickly than under the Gibbs sampler. 
Figures 2 and 4 in the Appendix demonstrate that running means based on Stan 
are in good agreement with those based on the Gibbs sampler.

\begin{figure}[htb]
\begin{center}
\includegraphics[height=5in,width=5in,angle=0]{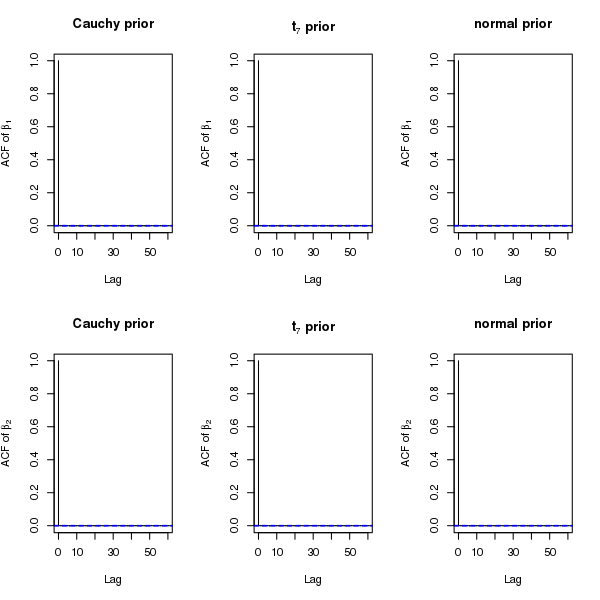}
\end{center}
\caption{Autocorrelation plots of the posterior samples of 
$\beta_1$ (top) and $\beta_2$ (bottom) from Stan, 
under independent Cauchy, $t_7$, and normal priors for the second simulated dataset.}
\label{fig:sim2acf:stan}
\end{figure}

The autocorrelation plots for Stan in Figure \ref{fig:sim2acf:stan} demonstrate a remarkable improvement 
over those for Gibbs in Figure \ref{fig:sim2acf} for all
priors, and the difference in mixing is the most prominent for Cauchy priors.

To summarize, all the plots unequivocally suggest that 
Cauchy priors lead to an extremely slow mixing Gibbs sampler 
and unusually large scales for the regression coefficients, 
even when all the marginal posterior means are guaranteed to exist. 
While mixing can be improved tremendously with Stan, the heavy tailed
posteriors under Stan are in agreement with those obtained from the Gibbs samplers.
One may argue that in spite of the unnaturally large regression coefficients, Cauchy priors 
could lead to superior predictions. 
Thus in the next two sections we compare predictions 
based on posteriors under the three priors for two real datasets. As Stan generates
nearly independent samples, we use Stan for MCMC simulations 
for the real datasets.

%
%

%
%

\section{Real Data}

\subsection{SPECT Dataset}
The ``SPECT'' dataset
\citep{Kurg:Cios:Tade:Ogie:Good:2001} is available from 
the UCI Machine Learning Repository\footnote{https://archive.ics.uci.edu/ml/datasets/SPECT+Heart}.
The binary response variable 
is whether a  patient's cardiac image is normal or abnormal, according to the diagnosis of cardiologists.
The predictors are 22 binary features obtained from the cardiac images using a machine learning algorithm. 
The goal of the study is to determine if
the predictors can correctly predict the diagnoses of cardiologists, 
so that the process could be automated to some extent.

Prior to centering, two of the
binary predictors are solitary quasicomplete separators: 
$x_{i, j}=0$  $\forall i \in A_{0}$ and $x_{i, j} \geq 0$ $\forall i \in A_{1}$, for $j=18, 19$,
with $\mathbf{X}_1$ denoting the column of ones.
\citet{Ghos:Reit:2013} analyzed this dataset with a Bayesian 
probit regression model which incorporated variable selection. 
As some of their proposed methods relied on an approximation 
of the marginal likelihood based on the MLE of $\boldsymbol{\beta}$, they had to
drop these potentially important predictors from the analysis. 
If one analyzed the dataset with the uncentered predictors, by Theorem
\ref{theorem:existence}, the posterior means $E(\beta_{18} \mid \mathbf{y})$ and $E(\beta_{19} \mid \mathbf{y})$ 
would not exist under  independent Cauchy priors.
However, after centering there are no solitary separators, so the posterior means of all coefficients exist.

The SPECT dataset is split into a training set of 80 observations and
a test set of 187 observations by \citet{Kurg:Cios:Tade:Ogie:Good:2001}. 
We use the former for model fitting and the latter for prediction.
First, for each of the three priors (Cauchy, $t_7$, and normal), 
we run Stan on the training dataset, for 1,000,000 iterations after discarding  
100,000 samples as burn-in. 

As in the simulation study, MCMC draws from Stan show excellent mixing for all
priors. However, the posterior means of the regression coefficients involved
in separation are rather large under Cauchy priors compared to the other priors.
For example, the posterior means of $(\beta_{18},\beta_{19})$ under
Cauchy, $t_7$, and normal priors are $(10.02,5.57), (3.24,1.68),$ and $(2.73,1.43)$ respectively.
These results suggest that Cauchy priors are too diffuse for datasets with separation.

Next for each $i=1,2,\dots, n_{\text{test}}$ in the test set, we estimate 
the corresponding success probability $\pi_{i}$ by the Monte Carlo average:
\begin{equation}
\widehat{\pi}_i^{\text{MC}} = \frac{1}{S}
	\sum_{s=1}^{S}\frac{e^{\mathbf{x}_i^{T} \boldsymbol{\beta}^{(s)}}}
	{1+e^{\mathbf{x}_i^{T} \boldsymbol{\beta}^{(s)}}},
\label{eqn:probmc}
\end{equation}
where $\boldsymbol{\beta}^{(s)}$ is the sampled value of 
$\boldsymbol{\beta}$ in iteration $s$, after burn-in. 
Recall that here $n_{\text{test}}=187$ and $S=10^6$. 
We calculate two different types of 
summary measures to assess predictive performance. 
We classify the $i$th observation in the test set as a success,
if $\widehat{\pi}_i^{\text{MC}}\geq 0.5$ and as a failure otherwise,
and compute the misclassification rates.
Note that the misclassification rate does not fully take into account 
the magnitude of $\widehat{\pi}_i^{\text{MC}}$.
For example, if $y_i=1$ both $\widehat{\pi}_i^{\text{MC}}=0.5$
and $\widehat{\pi}_i^{\text{MC}}=0.9$ would correctly classify the observation,
while the latter may be more preferable. 
So we also consider the average squared difference
between $y_i$ and $\widehat{\pi}_i^{\text{MC}}$:
\begin{equation}
MSE^{\text{MC}} = \frac{1}{n_{\text{test}}}
	\sum_{i=1}^{n_{\text{test}}}{(\widehat{\pi}_i^{\text{MC}}-y_i)}^2,
\label{eqn:msemc}
\end{equation}
which is always between 0 and 1, with a value closer to 0 being more preferable.
Note that the Brier score \citep{Brier_1950}
equals $2 MSE^{\text{MC}}$, according to its original definition.
Since in some modified definitions \citep{Blattenberger_Lad_1985}, 
it is the same as $MSE^{\text{MC}}$,  
we refer to $MSE^{\text{MC}}$ as the Brier score.

\begin{table}[ht]
\centering
\caption{Misclassification rates based on $\widehat{\pi}_i^{\text{MC}}$
and $\widehat{\pi}_i^{\text{EM}}$,
under Cauchy, $t_7$, and normal priors for the SPECT data. 
Small values are preferable.}
\label{tb:SPECT_cr}
\begin{tabular}{lccc}
  \hline
 & Cauchy & $t_7$ & normal \\ 
  \hline
MCMC & 0.273 & 0.257 & 0.251 \\ 
EM & 0.278 & 0.262 & 0.262 \\ 
   \hline
\end{tabular}
\end{table}

\begin{table}[ht]
\centering
\caption{Brier scores $MSE^{\text{MC}}$
and $MSE^{\text{EM}}$, under Cauchy, $t_7$, and normal priors  
for the SPECT data.  Small values are preferable.}
\label{tb:SPECT_mse}
\begin{tabular}{lccc}
  \hline
 & Cauchy & $t_7$ & normal \\ 
  \hline
MCMC & 0.172 & 0.165 & 0.163 \\ 
EM & 0.179 & 0.178 & 0.178 \\ 
   \hline
\end{tabular}
\end{table}

To compare the Monte Carlo estimates with those based on 
the EM algorithm of \citet{Gelman_etal_2008}, 
we also estimate the posterior mode, denoted by $\widetilde{\boldsymbol{\beta}}$ under identical priors and hyperparameters, using the 
R package {\tt arm} \citep{Gelman_etal_2015}. 
The EM estimator of $\pi_{i}$ is given by:
\begin{equation}
\widehat{\pi}_i^{\text{EM}}= \frac{e^{\mathbf{x}_i^{T} \widetilde{\boldsymbol{\beta}}}}
{1+e^{\mathbf{x}_i^{T} \widetilde{\boldsymbol{\beta}}}},
\label{eqn:probem}
\end{equation}
and $MSE^{\text{EM}}$ is calculated by replacing 
$\widehat{\pi}_i^{\text{MC}}$ by $\widehat{\pi}_i^{\text{EM}}$ in \eqref{eqn:msemc}.

We report the misclassification rates in 
Table \ref{tb:SPECT_cr} and the Brier scores in Table \ref{tb:SPECT_mse}.
MCMC achieves somewhat smaller misclassification rates and Brier scores than EM, 
especially under $t_{7}$ and normal priors.
This suggests that a full
Bayesian analysis using MCMC may produce estimates that are closer 
to the truth than modal estimates based on the EM algorithm.
The predictions are similar across the three prior distributions with 
the normal and $t_7$ priors yielding slightly more accurate results 
than Cauchy priors.


\subsection{Pima Indians Diabetes Dataset}
We now analyze the ``Pima Indians Diabetes'' dataset in the R package {\tt MASS}.
This is a classic dataset without separation that has been analyzed by many authors
in the past. Using this dataset we aim to compare predictions under different priors, when there is no 
separation. Using the training data provided in the package we predict the class labels of the test data. 
In this case the difference between different priors is practically nil. The Brier scores
are same up to three decimal places, across all priors and all methods (EM and MCMC). The
misclassification rates reported in Table \ref{tb:pima_cr} also show negligible
difference between priors and methods. Here Cauchy priors have a slightly better misclassification rate
compared to normal and $t_7$ priors, and MCMC provides slightly more accurate results compared to those
obtained from EM. These results suggest that when there is no separation and maximum likelihood estimates
exist, Cauchy priors may be preferable as default weakly informative
priors in the absence of real prior information.    

\begin{table}[ht]
\centering
\caption{Misclassification rates based on $\widehat{\pi}_i^{\text{MC}}$
and $\widehat{\pi}_i^{\text{EM}}$,
under Cauchy, $t_7$, and normal priors for the Pima Indians data. 
Small values are preferable.}
\label{tb:pima_cr}
\begin{tabular}{lccc}
  \hline
 & Cauchy & $t_7$ & normal \\ 
  \hline
MCMC & 0.196 & 0.199 & 0.199 \\ 
EM & 0.202 & 0.202 & 0.202 \\ 
   \hline
\end{tabular}
\end{table}

\section{Discussion}
We have proved that posterior means of regression coefficients in 
logistic regression are not always guaranteed to exist under 
the independent Cauchy priors recommended by \citet{Gelman_etal_2008}, 
if there is complete or quasicomplete separation in the data. 
In particular, we have introduced the notion of a solitary separator, 
which is a predictor capable of separating the samples on its own. 
Note that a solitary separator needs to be able to separate without the aid of 
any other predictor, not even the constant term corresponding to the intercept.
We have proved that for independent Cauchy priors,
the absence of solitary separators is 
a necessary condition for the existence of posterior means of all coefficients, 
for a general family of link functions in binary regression models. 
For logistic and probit regression, this has been shown to be 
a sufficient condition as well. In general, the sufficient condition 
depends on the form of the link function. 
We have also studied multivariate Cauchy priors,
where the solitary separator no longer plays an important role.
Instead, posterior means of all predictors exist if there is no separation, 
while none of them exist if there is complete separation. 
The result under quasicompelte separation is still unclear and will be
studied in future work.

In practice, after centering the input variables 
it is straightforward to check if there are solitary separators in the dataset.
The absence of solitary separators guarantees the existence of posterior means
of all regression coefficients in logistic regression under independent Cauchy priors.
However, our empirical results have shown that even when the posterior means
for Cauchy priors exist under separation, the posterior samples of the regression
coefficients may be extremely large in magnitude. Separation is usually
considered to be a sample phenomenon, so even if the predictors 
involved in separation are potentially important, some shrinkage 
of their coefficients is desirable through the prior. 
Our empirical results based on real datasets 
have demonstrated that the default Cauchy priors can lead to posterior means 
as large as 10, which is considered to be unusually large on the logit scale. 
Our impression is that Cauchy priors are good default choices in general because they contain weak prior information and let the data speak.
However, under separation, when there is little information in the data about the logistic regression coefficients (the MLE is not finite), 
it seems that lighter tailed priors, such as Student-$t$ priors with 
larger degrees of freedom or even normal priors, are more
desirable in terms of producing more plausible posterior distributions. 

From a computational perspective, we have observed very slow convergence of the Gibbs sampler
under Cauchy priors in the presence of separation. 
Note that if the design matrix is not of full column rank, for example when $p>n$, 
the $p$ columns of $\mathbf{X}$ will be linearly dependent. 
This implies that the equation for quasicomplete separation 
\eqref{eqn:qcompsep} will be satisfied with equality for all observations. 
Empirical results (not reported here for brevity) demonstrated that independent Cauchy priors
show convergence of the Gibbs sampler in this case also compared to other lighter tailed priors. 
Out-of-sample predictive performance based on a real dataset
with separation did not show the default Cauchy priors 
to be superior to $t_7$ or normal priors. 

In logistic regression, under a multivariate normal prior for 
$\boldsymbol{\beta}$, \citet{Choi_Hobert_2013} showed that 
the P{\'o}lya-Gamma data augmentation Gibbs sampler of 
\citet{Polson_etal_2013} is uniformly ergodic, and
the moment generating function of the posterior distribution
$p(\boldsymbol{\beta} \mid \mathbf{y})$ exists for all $\mathbf{X},\mathbf{y}$.
In our examples of datasets with separation, 
the normal priors led to the fastest convergence of the Gibbs sampler, 
reasonable scales for the posterior draws of $\boldsymbol{\beta}$, 
and comparable or even better
predictive performance than other priors. The results from Stan 
show no problem in mixing
under any of the priors. However, the problematic issue of posteriors with extremely 
heavy tails under Cauchy priors cannot be resolved without altering the prior.
Thus, after taking into account all the above considerations, 
for a full Bayesian analysis we recommend
the use of normal priors as a default, when there is separation. 
Alternatively, heavier tailed priors such as the $t_7$ could also be used 
if robustness is a concern.
On the other hand, if the goal of the analysis is to obtain point estimates 
rather than the entire posterior distribution, the posterior mode obtained from 
the EM algorithm of \citet{Gelman_etal_2015} under default Cauchy priors
\citep{Gelman_etal_2008} is a fast viable alternative.

\subsection*{Supplementary Material}
In the supplementary material, we present additional simulation
results for logistic and probit regression
with complete separation, along with an appendix that contains the proofs of all theoretical results.
The Gibbs sampler developed in the paper can be implemented  with the R package {\tt tglm}, available from the website: \url{https://cran.r-project.org/web/packages/tglm/index.html}.

\subsection*{Acknowledgement}
The authors thank the Editor-in-Chief, Editor, Associate Editor and
the reviewer for suggestions that led to a greatly improved paper.
The authors also thank Drs.~David Banks, James Berger,  William Bridges,
Merlise Clyde, Jon Forster, Jayanta Ghosh,  Aixin Tan, and Shouqiang Wang for helpful discussions. 
The research of Joyee Ghosh was partially supported by the NSA Young Investigator grant H98230-14-1-0126. 

\clearpage
\numberwithin{equation}{subsection}

\begin{center}
{\LARGE
\emph{
Supplementary Material: 
On the Use of Cauchy Prior Distributions for Bayesian Logistic Regression}}
\end{center}

\setcounter{section}{0}
\setcounter{figure}{0}

In this supplement, we first present an appendix with additional simulation results for 
logit and probit link functions, 
and then include an appendix that contains proofs of the theoretical results.

\section{Appendix: Simulation Results for Complete Separation Without Solitary Separators}
\label{section:sim2:app}
In this section we present some supporting figures for the analysis of
the simulated dataset described in Section \ref{section:sim2} 
of the manuscript under logit and probit links. 

\subsection{Logistic Regression for Complete Separation Without Solitary Separators}
\label{section:sim2:app:logit} 
\begin{figure}[h!]
\begin{center}
\includegraphics[height=5in,width=5in,angle=0]{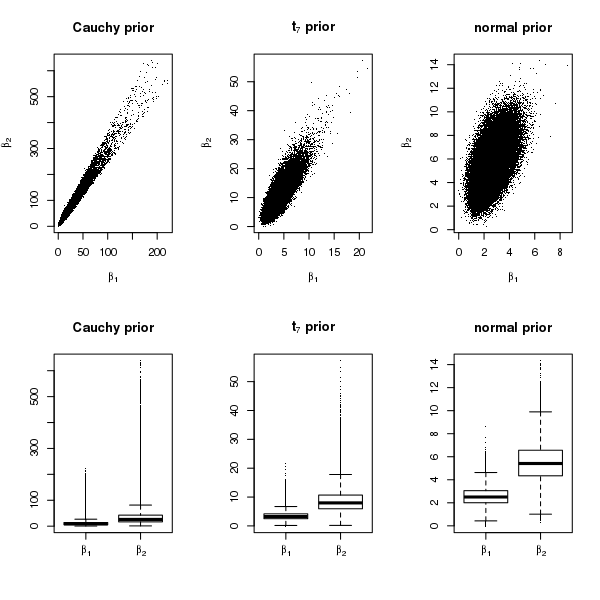}
\end{center}
\caption{Scatter plots (top) and box plots (bottom) of posterior samples  
of $\beta_1$ and $\beta_2$ for a logistic regression model, from the Gibbs sampler, under independent Cauchy, $t_7$, and normal priors 
for the second simulated dataset.}
\label{fig:app:sim2post}
\end{figure}

\begin{figure}[h!]
\begin{center}
\includegraphics[height=5in,width=5in,angle=0]{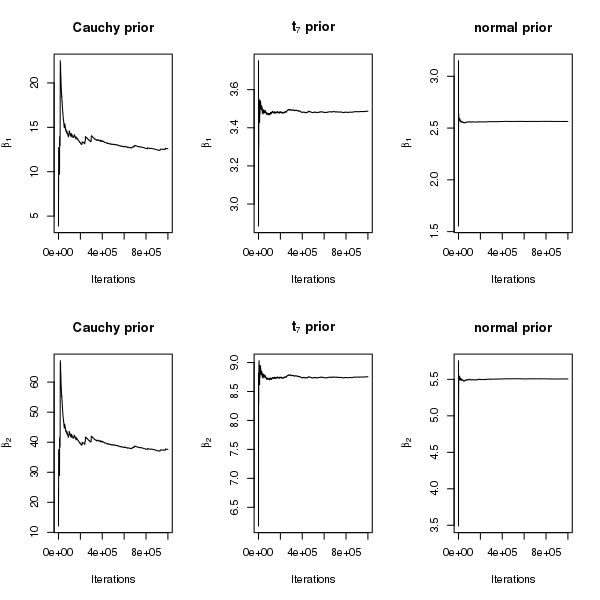}
\end{center}
\caption{Plots of running means of $\beta_1$ (top) and $\beta_2$
(bottom) sampled from the posterior distributions for a logistic regression model, via the Gibbs sampler, 
under independent Cauchy, $t_7$, and normal priors for the second simulated dataset. 
Here both $E(\beta_1 \mid \mathbf{y})$ and 
$E(\beta_2 \mid \mathbf{y})$ exist under independent Cauchy priors.}
\label{fig:app:sim2means}
\end{figure}

\begin{figure}[h!]
\begin{center}
\includegraphics[height=5in,width=5in,angle=0]{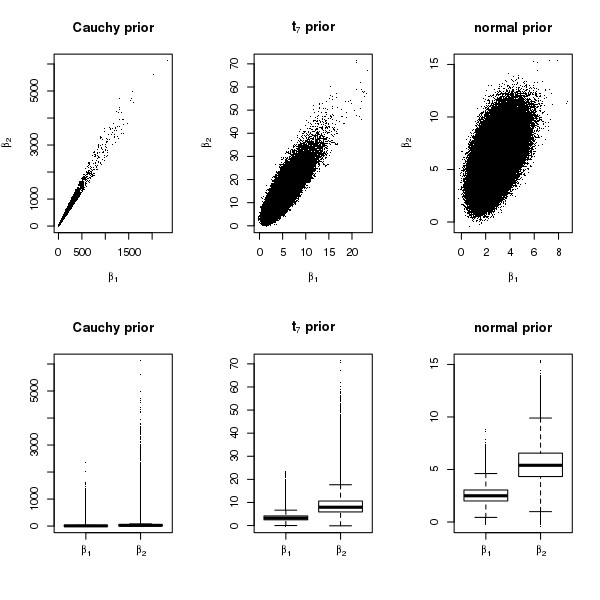}
\end{center}
\caption{Scatter plots (top) and box plots (bottom) of posterior samples  
of $\beta_1$ and $\beta_2$ for a logistic regression model, from Stan, under independent Cauchy, $t_7$, and normal priors 
for the second simulated dataset.}
\label{fig:app:sim2post:stan}
\end{figure}

\begin{figure}[h!]
\begin{center}
\includegraphics[height=5in,width=5in,angle=0]{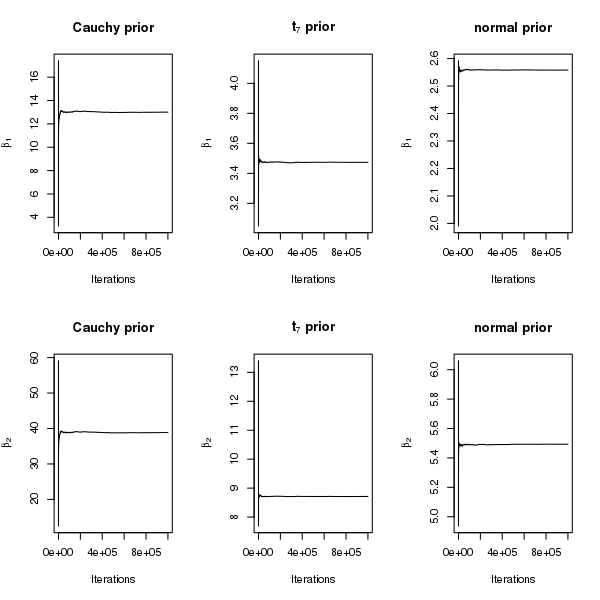}
\end{center}
\caption{Plots of running means of $\beta_1$ (top) and $\beta_2$
(bottom) sampled from the posterior distributions for a logistic regression model, via Stan,
under independent Cauchy, $t_7$, and normal priors for the second simulated dataset. 
Here both $E(\beta_1 \mid \mathbf{y})$ and 
$E(\beta_2 \mid \mathbf{y})$ exist under independent Cauchy priors.}
\label{fig:app:sim2means:stan}
\end{figure}

Figures \ref{fig:app:sim2post} and \ref{fig:app:sim2means} are based on the
posterior samples from a Gibbs sampler under a logit link, whereas 
Figures \ref{fig:app:sim2post:stan} and \ref{fig:app:sim2means:stan} are
corresponding results from Stan. A detailed discussion of the results
is provided in Section \ref{section:sim2} of the manuscript.

\subsection{Probit Regression for Complete Separation Without Solitary Separators}
\label{section:sim2probit}
\begin{figure}[h!]
\begin{center}
\includegraphics[height=5in,width=5in,angle=0]{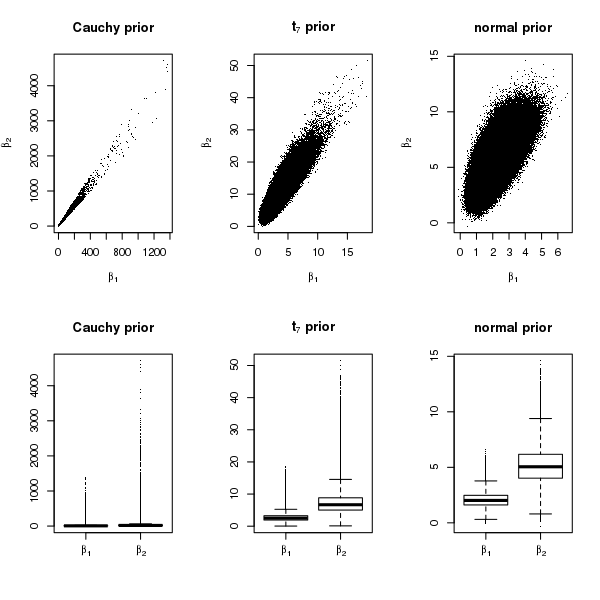}
\end{center}
\caption{Scatter plots (top) and box plots (bottom) of posterior samples  
of $\beta_1$ and $\beta_2$, for a probit regression model, under Cauchy, $t_7$, and normal priors 
for the second simulated dataset. Posterior sampling was generated via Stan.}
\label{fig:sim2post_probit}
\end{figure}

\begin{figure}[h!]
\begin{center}
\includegraphics[height=5in,width=5in,angle=0]{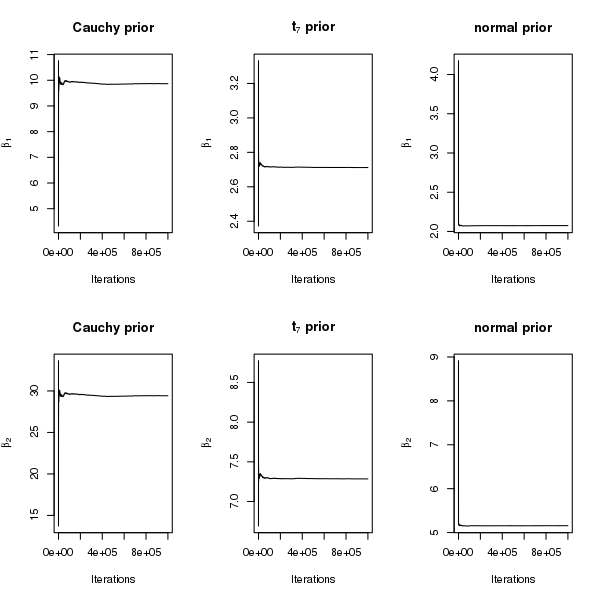}
\end{center}
\caption{Plots of running means of $\beta_1$ (top) and $\beta_2$
(bottom) sampled from the posterior distributions for a probit regression model,
under Cauchy, $t_7$, and normal priors for the second simulated datatset. 
Here both $E(\beta_1 \mid \mathbf{y})$ and 
$E(\beta_2 \mid \mathbf{y})$ exist under Cauchy priors. Posterior sampling was generated via Stan.}
\label{fig:sim2means_probit}
\end{figure}

In this section we analyze the simulated dataset described in Section \ref{section:sim2} 
of the manuscript under a probit link,
while keeping everything else the same. We have shown in Theorem \ref{theorem:general}, 
that the theoretical results hold for a probit link. 
The goal of this analysis is to demonstrate that the empirical results are similar under the logit and probit link functions.
For this dataset, Theorem \ref{theorem:general} guarantees the existence of both $E(\beta_1 \mid \mathbf{y})$ 
and $E(\beta_2 \mid \mathbf{y})$ under independent Cauchy priors and a probit link function.
As in the case of logistic regression the heavy tails of Cauchy priors translate into an extremely heavy right tail in the posterior distributions of $\beta_1$ and $\beta_2$, compared to the lighter tailed priors
(see Figure \ref{fig:sim2post_probit} and \ref{fig:sim2means_probit} here). Thus in the case of separation, 
normal priors seem to be reasonable for probit regression also.

\section{Appendix: Proofs}
\addcontentsline{toc}{section}{Appendices}
\renewcommand{\thesubsection}{\Alph{subsection}}
First, we decompose the proof of Theorem \ref{theorem:existence} into
two parts: in Appendix \ref{proof_th1_necessary} we show that 
a necessary condition for the existence of $E(\beta_j \mid \mathbf{y})$
is that $\mathbf{X}_j$ is not a solitary separator; and in
Appendix \ref{proof_th1_sufficient} we show that it is also a
sufficient condition. 
Then, we prove Theorem \ref{theorem:general} in
Appendix \ref{proof_th2}, and Corollary \ref{corollary:nonzero_prior_mean}
in Appendix \ref{proof_corollary:nonzero_prior_mean1}.
Finally, we decompose the proof of Theorem \ref{theorem:multivariate_Cauchy} 
into two parts: in Appendix \ref{proof_theorem:multivariate_Cauchy_no_separation}  
we show that all posterior means exist if there is no separation; then in Appendix
\ref{proof_theorem:multivariate_Cauchy_complete_separation} we show that 
none of the posterior means exist if there is complete separation. 

\subsection{Proof of the Necessary Condition for Theorem \ref{theorem:existence}}
\label{proof_th1_necessary}
Here we show that  if  $\mathbf{X}_j$ is a solitary separator, then
$E(\beta_j \mid \mathbf{y})$ does not exist, which is equivalent
to the necessary condition for Theorem \ref{theorem:existence}.
\begin{proof}
For notational simplicity, we define the functional form of the success and failure
probabilities in logistic regression as
\begin{equation}\label{eqn:f1f0}
 f_1(t) = e^t / (1 + e^t), \quad f_0(t) =1 -  f_1(t) = 1 / (1 + e^t),
\end{equation}
which are strictly increasing and decreasing functions of $t$, respectively.
In addition, both functions are bounded: $0 <  f_1(t),f_0(t) < 1$ for $t \in \mathbb{R}$.
Let $\boldsymbol\beta_{(-j)}$ and $\mathbf{x}_{i, (-j)}$ denote the vectors $\boldsymbol\beta$ and
$\mathbf{x}_i$ after excluding their $j$th entries $\beta_{j}$ and $x_{i,j}$, respectively.
Then the likelihood function can be written as
\begin{equation}\label{eq:likelihood}
	p(\mathbf{y} \mid \boldsymbol\beta)
	=  \prod_{i =1}^n p(y_i \mid  \boldsymbol\beta) 
	= \prod_{i\in A_1} f_1\left(x_{i,j}\beta_j + \mathbf{x}_{i, (-j)}^T \boldsymbol\beta_{(-j)}  \right) \cdot
\prod_{k \in A_0} f_0\left(x_{k,j}\beta_j + \mathbf{x}_{k, (-j)}^T \boldsymbol\beta_{(-j)} \right). 
\end{equation}
The posterior mean of $\beta_j$ exists provided $E(|\beta_j| \mid \mathbf{y})< \infty$.
When the posterior mean exists it is given by \eqref{eqn:pm_betaj}.
Clearly if one of the two integrals in \eqref{eqn:pm_betaj} is not finite, 
then $E(|\beta_j| \mid \mathbf{y}) = \infty$.
In this proof we will show that if $\alpha_j > 0$, the first integral in \eqref{eqn:pm_betaj} equals $\infty$. 
Similarly, it can be shown that if $\alpha_j < 0$, 
the second integral in \eqref{eqn:pm_betaj} equals $-\infty$. 

If $\alpha_j > 0$, by \eqref{eqn:compsep}-\eqref{eqn:solitary}, $x_{i,j} \geq 0$ for all $i \in A_1$, 
and $x_{k,j} \leq 0$ for all $k \in A_0$.
When $\beta_j > 0$, by the monotonic property of $f_1(t)$ and $f_0(t)$ we have, 
$p(\mathbf{y} \mid \boldsymbol\beta) \geq
\prod_{i\in A_1} f_1\left(\mathbf{x}_{i, (-j)}^T \boldsymbol\beta_{(-j)}  \right)
\cdot \prod_{k \in A_0} f_0\left(\mathbf{x}_{k, (-j)}^T \boldsymbol\beta_{(-j)} \right)$,
which is free of $\beta_j$. Therefore,
\begin{align} 
&	\int_0^{\infty} \beta_j ~p(\beta_j \mid \mathbf{y})~d\beta_j  
= 	 \int_0^{\infty} \beta_j \left[ \int_{\mathbb{R}^{p-1}} 
	\frac{p(\mathbf{y} \mid \boldsymbol\beta) p(\beta_j) p(\boldsymbol\beta_{(-j)})}{p(\mathbf{y})} 
	d\boldsymbol\beta_{(-j)}\right] ~d\beta_j  \nonumber \\
=~&	 \frac{1}{p(\mathbf{y})}\int_0^{\infty} \beta_j p(\beta_j)  \left[ \int_{\mathbb{R}^{p-1}} 
	p(\mathbf{y} \mid \boldsymbol\beta) p(\boldsymbol\beta_{(-j)})
	d\boldsymbol\beta_{(-j)}\right] ~d\beta_j   \nonumber  \\
\geq ~&  \frac{1}{p(\mathbf{y})}\int_0^{\infty} \beta_j p(\beta_j) \left[ \int_{\mathbb{R}^{p-1}} \prod_{i\in A_1} f_1\left(\mathbf{x}_{i, (-j)}^T \boldsymbol\beta_{(-j)}  \right)
	 \prod_{k \in A_0} f_0\left(\mathbf{x}_{k, (-j)}^T \boldsymbol\beta_{(-j)} \right)
	p(\boldsymbol\beta_{(-j)})
	d\boldsymbol\beta_{(-j)}\right] ~d\beta_j \nonumber \\
= ~ &	 \frac{\int_0^{\infty} \beta_j p(\beta_j)~d\beta_j}{p(\mathbf{y})}
	\left[ \int_{\mathbb{R}^{p-1}} 
	\prod_{i\in A_1} f_1\left(\mathbf{x}_{i, (-j)}^T \boldsymbol\beta_{(-j)}  \right)
	 \prod_{k \in A_0} f_0\left(\mathbf{x}_{k, (-j)}^T \boldsymbol\beta_{(-j)} \right)
	p(\boldsymbol\beta_{(-j)})
	d\boldsymbol\beta_{(-j)}\right]. 
 \label{eqn:pm_betaj_plus} 
\end{align}
Here the first equation results from independent priors, i.e., 
$p(\bm{\beta}_{(-j)} \mid \beta_j) = p(\bm{\beta}_{(-j)})$.
Since $ p(\mathbf{y} \mid \boldsymbol\beta) < 1$,
$p(\mathbf{y}) = \int_{\mathbb{R}^p} p(\mathbf{y} \mid \boldsymbol\beta) 
p(\boldsymbol\beta) d\boldsymbol\beta < \int_{\mathbb{R}^p}
p(\boldsymbol\beta) d\boldsymbol\beta=1$. 
Moreover, $p(\mathbf{y} \mid \boldsymbol\beta)
p(\boldsymbol\beta)> 0$ for all $\boldsymbol\beta \in \mathbb{R}^p$, so
we also have $p(\mathbf{y}) > 0$, implying that $0<p(\mathbf{y}) <1$.   
For the independent Cauchy priors in \eqref{eqn:prior}, $\int_0^{\infty} \beta_j
p(\beta_j) ~d\beta_j = \infty$ and the second integral in \eqref{eqn:pm_betaj_plus} is positive, 
hence \eqref{eqn:pm_betaj_plus} equals $\infty$.
\end{proof}

\subsection{Proof of the Sufficient Condition for Theorem \ref{theorem:existence}}\label{proof_th1_sufficient}
Here we show that if $\mathbf{X}_j$ is not a solitary separator, 
then the posterior mean of $\beta_j$ exists. 
\begin{proof}
When $E(|\beta_j| \mid \mathbf{y})< \infty$ the posterior mean
  of $\beta_j$ exists and is given by
\begin{align}\nonumber
E(\beta_j \mid \mathbf{y})
&	= \int_{-\infty}^{\infty} \beta_j p(\beta_j \mid \mathbf{y}) d \beta_j\\ \label{eq:posterior_mean_beta_j}
&	=  \frac{1}{p(\mathbf{y})}
	\underbrace{  \int_{0}^{\infty} \beta_j p(\beta_j) p(\mathbf{y} \mid \beta_j) d \beta_j
	}_{\text{denoted by } E(\beta_j \mid \mathbf{y})^+}
	+ 
	\frac{1}{p(\mathbf{y})}
	\underbrace{  \int_{-\infty}^{0} \beta_j p(\beta_j) p(\mathbf{y} \mid \beta_j) d \beta_j
	}_{\text{denoted by } E(\beta_j \mid \mathbf{y})^-}.
\end{align}
Since $0<p(\mathbf{y})<1$ it is enough to show that 
the positive term $E(\beta_j \mid \mathbf{y})^+$ has a finite upper bound, and 
the negative term $E(\beta_j \mid \mathbf{y})^-$ has a finite lower bound.
For notational simplicity, in the remainder of the proof, we let
$\bm{\alpha}_{(-j)}$, $ \mathbf{x}_{i, (-j)}$, $\bm{\beta}_{(-j)}$, and $\bm{\sigma}_{(-j)}$
denote the vectors $\bm{\alpha}$, $\mathbf{x}_i$, $\bm{\beta}$, 
and $\bm{\sigma} = (\sigma_1, \ldots, \sigma_p)^T$ 
after excluding their $j$th entries, respectively.

{\bf We first show that $E(\beta_j \mid \mathbf{y})^+$ has a finite upper bound.}
Because $\mathbf{X}_j$ is not a solitary separator, 
there exists either
\begin{enumerate}[label=(\alph*)]
\item \label{condition_a} an $i' \in A_1$ such that $x_{i', j} < 0$, or
\item \label{condition_b} a $k' \in A_0$, such that $x_{k', j} > 0$.
\end{enumerate}
\noindent If both \ref{condition_a} and \ref{condition_b} are violated then $\mathbf{X}_j$ is a solitary
separator and leads to a contradiction.
Furthermore, for such $\mathbf{x}_{i', (-j)}$ or $\mathbf{x}_{k', (-j)}$, it contains
at least one non-zero entry.
This is because if $j \neq 1$, i.e., $\mathbf{X}_j$ does not
correspond to the intercept (the column of all one's), 
then the first entry in $\mathbf{x}_{i', (-j)}$ or $\mathbf{x}_{k', (-j)}$ equals $1$.
If $j = 1$, due to the assumption that $\mathbf{X}$ has a column
rank $>1$, there exists one row $i^{\diamond} \in \{1, 2, \ldots, n\}$ such that $\mathbf{x}_{i^{\diamond}, (-1)}$ 
contains at least one non-zero entry.
If $i^{\diamond} \in A_0$, then we let $k' = i^{\diamond}$ and the
condition \ref{condition_b} holds because $x_{i^{\diamond}, 1} = 1 > 0$. 
If $i^{\diamond} \in A_1$, we may first rescale $\mathbf{X}_1$ by $-1$,
which transforms $\beta_1$ to $-\beta_1$. Since the Cauchy prior centered at zero
is invariant to this rescaling, and $E(-\beta_1 \mid \mathbf{y})$
exists if and only if $E(\beta_1 \mid \mathbf{y})$ exists,
we can just apply this rescaling, after which $x_{i^{\diamond}, 1} = -1 < 0$.
Then we let $i' = i^{\diamond}$ and the condition  \ref{condition_a} holds.

We first assume that condition \ref{condition_a} is true. 
We define a positive constant $\epsilon = |x_{i',j}| / 2 = -x_{i',j} / 2$.
For any $\beta_j > 0$, we define a subset of the domain of $\bm{\beta}_{(-j)} $
as follows
\begin{equation}\label{eq:G}
G(\beta_j) \stackrel{\text{def}}{=} \left\{  
\bm{\beta}_{(-j)}  \in \mathbb{R}^{p-1}: \mathbf{x}_{i', (-j)}^T \bm{\beta}_{(-j)} 
< \epsilon \beta_j \right\}.
\end{equation}
Then for any $\bm{\beta}_{(-j)} \in G(\beta_j)$, 
$\mathbf{x}_{i'}^T \bm{\beta} 
= x_{i',j}\beta_j+ \mathbf{x}_{i', (-j)}^T \bm{\beta}_{(-j)} 
<(x_{i',j} + \epsilon)  \beta_j= - \epsilon  \beta_j$.
Therefore, 
\begin{equation}\label{eqn:G}
f_1(\mathbf{x}_{i'}^T \bm{\beta} ) < f_1(- \epsilon  \beta_j), \quad \text{for all } \bm{\beta}_{(-j)} \in G(\beta_j).
\end{equation}
As $f_1(\cdot)$ and $f_0(\cdot)$ are bounded above by 1, 
the likelihood function $p(\mathbf{y} \mid \bm{\beta})$ in \eqref{eq:likelihood}
is bounded above by $f_1(\mathbf{x}_{i'}^T \bm{\beta})$. Thus  
\begin{align} 
&	E(\beta_j \mid \mathbf{y})^+
	= \int_{0}^{\infty} \beta_j p(\beta_j) 
	\left[ \int_{\mathbb{R}^{p-1}} p(\mathbf{y} \mid \bm{\beta}) 
	p(\bm{\beta}_{(-j)}) d\bm{\beta}_{(-j)} \right] d \beta_j \nonumber \\
<~&	\int_{0}^{\infty} \beta_j p(\beta_j) 
	\left[ \int_{\mathbb{R}^{p-1}} f_1(\mathbf{x}_{i'}^T \bm{\beta}) 
	p(\bm{\beta}_{(-j)}) d\bm{\beta}_{(-j)} \right] d \beta_j \nonumber \\
=~&	\int_{0}^{\infty} \beta_j p(\beta_j) 
	\left[ \int_{G(\beta_j)} f_1(\mathbf{x}_{i'}^T \bm{\beta}) 
	p(\bm{\beta}_{(-j)}) d\bm{\beta}_{(-j)} 
	+ \int_{\mathbb{R}^{p-1}\backslash G(\beta_j)} f_1(\mathbf{x}_{i'}^T \bm{\beta})
	p(\bm{\beta}_{(-j)}) d\bm{\beta}_{(-j)} \right] d \beta_j\nonumber \\
<~&	\int_{0}^{\infty} \beta_j p(\beta_j) 
	\left[ \int_{G(\beta_j)} f_1(- \epsilon \beta_j) 
	p(\bm{\beta}_{(-j)}) d\bm{\beta}_{(-j)} 
	+ \int_{\mathbb{R}^{p-1}\backslash G(\beta_j)}  
	p(\bm{\beta}_{(-j)}) d\bm{\beta}_{(-j)} \right] d \beta_j \label{eqn:pm_gamma1_plus}
\end{align}
Here the last inequality results from \eqref{eqn:G} and the fact that 
the function $f_1(\cdot)$ is bounded above by $1$.
An upper bound is obtained for the first term in the bracket in \eqref{eqn:pm_gamma1_plus}
using the fact that the integrand is non-negative as follows:
\begin{align} \nonumber
\int_{G(\beta_j)} f_1(- \epsilon \beta_j) 
	p(\bm{\beta}_{(-j)}) d\bm{\beta}_{(-j)} 
&< 	\int_{\mathbb{R}^{p-1}} f_1(- \epsilon \beta_j) 
	p(\bm{\beta}_{(-j)}) d\bm{\beta}_{(-j)} \\ \label{eqn:pm_gamma1_plus3}
&=	f_1(- \epsilon \beta_j) 
\underbrace{\int_{\mathbb{R}^{p-1}}  
	p(\bm{\beta}_{(-j)}) d\bm{\beta}_{(-j)}}_{=1} 
=	\frac{e^{-\epsilon \beta_j}}{1 + e^{-\epsilon \beta_j}}
< 	e^{-\epsilon \beta_j}. 
\end{align}
Recall that $\mathbf{x}_{i', (-j)}$ contains at least one non-zero entry. 
We assume that $\mathbf{x}_{i', r} \neq 0$. Then
to simplify the second term in the bracket in \eqref{eqn:pm_gamma1_plus},
we transform $\bm{\beta}_{(-j)}$ to $(\eta, \bm{\xi})$ via a linear transformation, such that
$\eta = \mathbf{x}_{i', (-j)}^T \bm{\beta}_{(-j)}$, and $\bm{\xi}$ is 
the vector $\bm{\beta}_{(-j)}$ after excluding $\beta_r$.  
The characteristic function of a Cauchy distribution
  $C(\mu,\sigma)$ is $\varphi(t) = e^{it\mu - |t|\sigma}$, where $t \in \mathbb{R}$.
Since {\it a priori}, $\eta$ is a linear combination of independent $C(0,\sigma_{\ell})$ random variables,
$\beta_\ell$, for $1\leq \ell \leq p, \ell \neq j$, 
its characteristic function is
\[
\varphi_\eta(t) = E(e^{it \eta}) 
= \prod_{1\leq \ell \leq p, \ell \neq j} E\left[e^{i(t x_{i',\ell}) \beta_\ell}\right]
= \prod_{1\leq \ell \leq p, \ell \neq j} \varphi_{\beta_\ell}(t x_{i', \ell}) 
= e^{ - |t| \sum_{1\leq \ell \leq p, \ell \neq j} |x_{i', \ell}| \sigma_\ell}.
\]
So the induced prior of $\eta$ is $C(0, \sum_{1\leq \ell \leq p, \ell \neq j} |x_{i', \ell}| \sigma_\ell)$.
Let $|\mathbf{x}_{i', (-j)}|$ denote the vector obtained by taking absolute values of each element
of $\mathbf{x}_{i', (-j)}$, then the above scale parameter  
$\sum_{1\leq \ell \leq p, \ell \neq j} |x_{i', \ell}| \sigma_\ell$= $|\mathbf{x}_{i', (-j)}|^T \bm{\sigma}_{(-j)}$.
By \eqref{eq:G}, for any $\bm{\beta}_{(-j)} \not\in G(\beta_j)$, 
the corresponding $\eta \geq \epsilon \beta_j$
and $\bm{\xi} \in \mathbb{R}^{p-2}$.
An upper bound is calculated for the second term in the bracket in \eqref{eqn:pm_gamma1_plus}.
Note that $p(\eta)p(\bm{\xi} \mid \eta)$ is the joint density of $\eta$ and $\bm{\xi}$. Since it 
incorporates the Jacobian of transformation from $\bm{\beta}_{(-j)} $
to $\eta$ and $\bm{\xi}$, a separate Jacobian term is
not needed in the first equality below.
\begin{align} \nonumber
\int_{\mathbb{R}^{p-1}\backslash G(\beta_j)}  
	p(\bm{\beta}_{(-j)}) d\bm{\beta}_{(-j)} 
&= 	\int_{\epsilon\beta_j}^{\infty}  \int_{\mathbb{R}^{p-2}}
	p(\eta)p(\bm{\xi} \mid \eta) d \bm{\xi} d \eta \\ \nonumber
&=	\int_{\epsilon\beta_j}^{\infty}  
	\frac{\int_{\mathbb{R}^{p-2}} p(\bm{\xi} \mid \eta) d \bm{\xi} }
	{\pi~ |\mathbf{x}_{i', (-j)}|^T \bm{\sigma}_{(-j)}
	\left[ 1 + \eta^2 / 
	\left(|\mathbf{x}_{i', (-j)}|^T \bm{\sigma}_{(-j)}\right)^2
      \right]} d \eta\\ \nonumber
&=	\frac{1}{\pi}\left[\frac{\pi}{2} - \arctan \left( \frac{\epsilon\beta_j}
	{|\mathbf{x}_{i', (-j)}|^T \bm{\sigma}_{(-j)}} \right)  \right]\\ 
&=	\frac{1}{\pi}\arctan \left( \frac{|\mathbf{x}_{i', (-j)}|^T \bm{\sigma}_{(-j)}}{\epsilon\beta_j}\right) 
<	\frac{|\mathbf{x}_{i', (-j)}|^T
  \bm{\sigma}_{(-j)}}{\pi\epsilon\beta_j}.
\label{eqn:upperbound2}
\end{align}
Here, the second equality holds because $\int_{\mathbb{R}^{p-2}} p(\bm{\xi} \mid \eta) d \bm{\xi} = 1$;
the last inequality holds because $\epsilon$ and $\beta_j$ are both positive,
and for any $t > 0$, $\arctan(t) < t$.

Then substituting the expression for $p(\beta_j)$ as in \eqref{eqn:prior},
we continue with \eqref{eqn:pm_gamma1_plus} to find an upper bound. 
\begin{align} 
E(\beta_j \mid \mathbf{y})^+
&	< \int_{0}^{\infty} \frac{\beta_j}{\pi \sigma_j (1 +  \beta_j^2 / \sigma_j^2)} \left[ e^{-\epsilon \beta_j} 
	+ \frac{|\mathbf{x}_{i', (-j)}|^T
          \bm{\sigma}_{(-j)}}{\pi\epsilon\beta_j} \right] d \beta_j
      \nonumber \\
&	< \int_{0}^{\infty} \frac{\beta_j e^{-\epsilon \beta_j}}{\pi \sigma_j } d \beta_j
	+ \int_{0}^{\infty} \frac{|\mathbf{x}_{i', (-j)}|^T \bm{\sigma}_{(-j)}}
	{\pi^2 \sigma_j \epsilon (1 +  \beta_j^2 / \sigma_j^2)} d \beta_j
= 	\frac{1}{\pi \sigma_j \epsilon^2} + \frac{|\mathbf{x}_{i',
    (-j)}|^T \bm{\sigma}_{(-j)}}{2\pi\epsilon}. 
\label{eqn:upperbound+}
\end{align}
On the other hand, if condition \ref{condition_b} holds, then we just need to slightly modify the above proof.
We define $\epsilon = |x_{k', j}|/2 = x_{k', j}/2$, and change \eqref{eq:G}
to 
\[
G(\beta_j) \stackrel{\text{def}}{=} \left\{  
\bm{\beta}_{(-j)}  \in \mathbb{R}^{p-1}: \mathbf{x}_{k', (-j)}^T \bm{\beta}_{(-j)} 
> - \epsilon \beta_j \right\}.
\] 
Consequently, the terms 
$f_1(\mathbf{x}_{i'}^T \bm{\beta})$ and $f_1(- \epsilon \beta_j)$
in \eqref{eqn:pm_gamma1_plus} have to be changed to 
$f_0(\mathbf{x}_{k'}^T \bm{\beta})$ and $f_0(\epsilon \beta_j)$, respectively.
For the logit link, $f_0(\epsilon \beta_j) = f_1(- \epsilon \beta_j)$.
The range of the integral in \eqref{eqn:upperbound2} with respect to $\eta$
is from $-\infty$ to $-\epsilon\beta_j$;
however, because the density of $\eta$ is symmetric around 0, the
 value of the integral stays the same. So it can be shown an upper bound for $E(\beta_j \mid \mathbf{y})^+$ is 
$\left[1 / \left(\pi \sigma_j \epsilon^2\right)\right] + \left[|\mathbf{x}_{k',(-j)}|^T
\bm{\sigma}_{(-j)}/ \left(2\pi\epsilon \right)\right]$.

{\bf We now show that the negative term $E(\beta_j \mid \mathbf{y})^-$ has a finite lower bound. }
For any $\beta_j < 0$, by expressing $\beta_j^* = - \beta_j$, we need to show that
the positive term
$-E(\beta_j \mid \mathbf{y})^- = - \int_{-\infty}^{0} \beta_j p(\beta_j) p(\mathbf{y} \mid \beta_j) d \beta_j
= \int_0^{\infty} \beta_j^* p(\beta_j^*) p(\mathbf{y} \mid -\beta_j^*) d \beta_j^*$
has a finite upper bound. As the idea is very similar to the proof 
of existence of $E(\beta_j \mid \mathbf{y})^+$, we present less
details here.

Since the predictor $\mathbf{X}_j$ is not a solitary separator, 
 there exists either 
\begin{enumerate}[label=(\alph*), start = 3]
\item \label{condition_c} an  $i^* \in A_1$ such that $x_{i^*, j} > 0$, or
\item \label{condition_d} a  $k^* \in A_0$, such that $x_{k^*, j} < 0$.
\end{enumerate}

\noindent If \ref{condition_c} and \ref{condition_d} are
both violated $\mathbf{X}_j$ has to be a solitary separator, which 
leads to a contradiction.
WLOG, we assume that condition \ref{condition_c} is true, and as before
$\mathbf{x}_{i^*, (-j)}$ must contain at least one non-zero entry,
say, $\mathbf{x}_{i^*, s} \neq 0$.
If condition \ref{condition_d} is true, then we can adopt a 
modification similar to the one that is used to prove the existence under condition 
\ref{condition_b} based on the proof under condition \ref{condition_a}.

We define a positive constant $\epsilon = x_{i^*, j}/2$. For any $\beta_j^* > 0$, we define
a subset of $\mathbb{R}^{p-1}$ as
$G(\beta_j^*)\stackrel{\text{def}}{=} \left\{  
\bm{\beta}_{(-j)}  \in \mathbb{R}^{p-1}: \mathbf{x}_{i^*, (-j)}^T \bm{\beta}_{(-j)} 
< \epsilon \beta_j^* \right\}$.
Then for all $\bm{\beta}_{(-j)} \in G(\beta_j^*)$, 
$\mathbf{x}_{i^*}^T\bm{\beta} = -x_{i^*, j}\beta_j^* + \mathbf{x}_{i^*, (-j)}^T \bm{\beta}_{(-j)} 
< (-x_{i^*,j} + \epsilon)  \beta_j^* = - \epsilon \beta_j^*$, hence
$f_1(\mathbf{x}_{i^*}^T\bm{\beta}) < f_1\left(- \epsilon \beta_j^* \right)$.
Since the likelihood function 
$p(\mathbf{y} \mid -\beta_j^*, \bm{\beta}_{(-j)})  <
f_1(x_{i^*, j}(-\beta_j^*) + \mathbf{x}_{i^*, (-j)}^T \bm{\beta}_{(-j)} ) < 1$, 
\begin{align} 
&	-E(\beta_j \mid \mathbf{y})^-
	= \int_{0}^{\infty} \beta_j^* p(\beta_j^*) 
	\left[ \int_{\mathbb{R}^{p-1}} p(\mathbf{y} \mid -\beta_j^*, \bm{\beta}_{(-j)})
	p(\bm{\beta}_{(-j)}) d\bm{\beta}_{(-j)} \right] d \beta_j^* \nonumber \\
<~&	\int_{0}^{\infty} \beta_j^* p(\beta_j^*) 
	\left[ \int_{G(\beta_j^*)} f_1(- \epsilon \beta_j^*) 
	p(\bm{\beta}_{(-j)}) d\bm{\beta}_{(-j)} 
	+ \int_{\mathbb{R}^{p-1}\backslash G(\beta_j^*)}  
	p(\bm{\beta}_{(-j)}) d\bm{\beta}_{(-j)} \right] d \beta_j^*. \label{eqn:pm_gamma1_plus2}
\end{align}
The first term in the bracket in \eqref{eqn:pm_gamma1_plus2} has an upper bound
$\exp(-\epsilon \beta_j^*)$ as in \eqref{eqn:pm_gamma1_plus3}.
Recall that $\mathbf{x}_{i^*, s} \neq 0$. 
We now transform $\bm{\beta}_{(-j)}$ to $(\eta, \bm{\xi})$ via 
a linear transformation, such that
$\eta = \mathbf{x}_{i^*, (-j)}^T \bm{\beta}_{(-j)}$ and $\bm{\xi}$ is 
the vector $\bm{\beta}_{(-j)}$ after excluding $\beta_s$.
The prior of $\eta$ is  $C(0,|\mathbf{x}_{i^*, (-j)}|^T
\bm{\sigma}_{(-j)})$. For any $\bm{\beta}_{(-j)} \not\in G(\beta_j^*)$, the corresponding $\eta \geq \epsilon \beta_j^*$.
Therefore as in \eqref{eqn:upperbound2}, we obtain an upper bound for the second term in
the bracket in \eqref{eqn:pm_gamma1_plus2} as 
$|\mathbf{x}_{i^*, (-j)}|^T \bm{\sigma}_{(-j)}/\left(\pi\epsilon\beta_j^*\right)$.
Finally, following \eqref{eqn:upperbound+} an upper bound for
$-E(\beta_j \mid \mathbf{y})^-$ is
$\left[1/\left(\pi \sigma_j \epsilon^2\right)\right] + \left[|\mathbf{x}_{i^*, (-j)}|^T \bm{\sigma}_{(-j)}/ \left(2\pi\epsilon\right)\right]$.
\end{proof}

\subsection{Proof of Theorem \ref{theorem:general}}\label{proof_th2}
\begin{proof}
Following the proof of Theorem \ref{theorem:existence}, 
we denote the success probability function by $\pi = f_1(\mathbf{x}^T \bm{\beta})$,
where $f_1(\cdot)$ is the inverse link function, i.e., $f_1(\cdot) = g^{-1}(\cdot)$.
Similarly, let the failure probability function be $f_0(\mathbf{x}^T \bm{\beta}) = 1 - f_1(\mathbf{x}^T \bm{\beta})$.
Note that the proof of the necessary condition for Theorem \ref{theorem:existence} 
given in Appendix \ref{proof_th1_necessary} 
only relies on the fact the $f_1(\cdot)$ is increasing, continuous,
and bounded between $0$ and $1$.
Since the link function $g(\cdot)$ is assumed to be strictly increasing and differentiable,
so is $f_1(\cdot)$. Moreover, the range of $f_1$ is $(0, 1)$.
Therefore, the proof of the necessary condition for Theorem \ref{theorem:general} follows immediately.

For the proof of the sufficient condition in Theorem \ref{theorem:general}, one can follow the
 proof in Appendix \ref{proof_th1_sufficient} and proceed with the
 specific choice of $\epsilon$ used there, when condition \ref{condition_a}
 holds. The proof is identical until
 \eqref{eqn:pm_gamma1_plus} because the
 explicit form of the inverse link function is not used until that step. We
 re-write the final step in \eqref{eqn:pm_gamma1_plus} below and
 proceed from there:
\begin{align}\nonumber
&	E(\beta_j \mid \mathbf{y})^+ \\ 
 <&	\int_{0}^{\infty} \beta_j p(\beta_j) 
	\left[ \int_{G(\beta_j)} f_1(- \epsilon \beta_j) 
	p(\bm{\beta}_{(-j)}) d\bm{\beta}_{(-j)}  
+  \int_{\mathbb{R}^{p-1}\backslash G(\beta_j)}  
	p(\bm{\beta}_{(-j)}) d\bm{\beta}_{(-j)} \right] d \beta_j
      \nonumber \\
= & \int_{0}^{\infty} \beta_j p(\beta_j) f_1(- \epsilon \beta_j) 
	\underbrace{\left[ \int_{G(\beta_j)}p(\bm{\beta}_{(-j)}) d\bm{\beta}_{(-j)} \right]
	}_{<  \int_{\mathbb{R}^{p-1}}p(\bm{\beta}_{(-j)}) d\bm{\beta}_{(-j)} =1} d \beta_j 
	 \nonumber \\
& + \int_{0}^{\infty} \beta_j p(\beta_j) \left[ \int_{\mathbb{R}^{p-1}\backslash G(\beta_j)}  
	p(\bm{\beta}_{(-j)}) d\bm{\beta}_{(-j)} \right] d \beta_j 
      \nonumber \\
<&  \int_{0}^{\infty} \beta_j p(\beta_j) \underbrace{f_1(- \epsilon \beta_j)}_{=g^{-1}(- \epsilon \beta_j)}d \beta_j 
	+ \int_{0}^{\infty} \beta_j p(\beta_j) \left[ \int_{\mathbb{R}^{p-1}\backslash G(\beta_j)}  
	p(\bm{\beta}_{(-j)}) d\bm{\beta}_{(-j)} \right] d \beta_j.
\label{eqn:th2:sufficient:proof}
\end{align} 
The sufficient condition in Theorem \ref{theorem:general} states that
for every positive constant $\epsilon$, 
\[
\int_{0}^{\infty} \beta_j
p(\beta_j) g^{-1}(- \epsilon \beta_j) d \beta_j < \infty.
\]
This implies
that the integral will be bounded for the specific choice of $\epsilon$ used in the
above proof, and hence the first integral in
\eqref{eqn:th2:sufficient:proof} is bounded above. The second integral
does not depend on the link function and its bound can be obtained
exactly as in Appendix \ref{proof_th1_sufficient}. Thus $E(\beta_j \mid
\mathbf{y})^+ < \infty$ under condition (a). On the other hand if 
condition (b) holds, the proof follows
similarly as in Appendix \ref{proof_th1_sufficient}  and now we 
need to use the sufficient condition in Theorem \ref{theorem:general} that
for every positive constant $\epsilon$, $\int_{0}^{\infty} \beta_j p(\beta_j)
\left[1-g^{-1}(\epsilon \beta_j)\right] d \beta_j< \infty$.
A bound for $-E(\beta_j \mid \mathbf{y})^-$
can be obtained similarly, which completes the proof.

In probit regression, we first show that
\begin{equation}\label{eq:probit_logit}
g^{-1}_{\text{probit}}(t) = \Phi(t)
< e^t/(1 + e^t) = g^{-1}_{\text{logit}}(t),\quad \text{for any } t < 0, 
\end{equation}
where $\Phi(t)$ is the standard normal cdf.
It is equivalent to show that the difference function 
\begin{equation}\label{eq:u}
u(t) = \Phi(t) - \frac{e^t}{1 + e^t} < 0, \text{ for all } t < 0.
\end{equation}
Note that $u(0) = 1/2 - 1/2 = 0$, and $\lim_{t\rightarrow -\infty} u(t) = 0$.
Since $u(t)$ is differentiable, 
we have
\[
u'(t) = \frac{1}{\sqrt{2\pi}} e^{-\frac{t^2}{2}} - \frac{e^t}{(1 + e^t)^2}.\]
Now $u'(0) = 1/\sqrt{2\pi} - 1/4 >0$, and 
when $t$ is very small, $u'(t) < 0$ since $e^{-t^2/2}$ decays to zero at 
a faster speed than $e^t$, i.e., there exists a $\tilde{t}<0$ such that $u'(\tilde{t})<0$. Since $u'(t)$ is 
a continuous function, the intermediate value theorem applied to $[\tilde{t},0]$ shows that 
there exists a $t_1 < 0$ such that $u'(t_1) = 0$.
Therefore, to show \eqref{eq:u}, it is sufficient to show that
$u'(t)$ has a unique root on $\mathbb{R}^-$, which is proved by contradiction as follows.

If $u'(t)$ has two distinct roots $t_1, t_2 < 0$, i.e., for $i = 1, 2$,
$u'(t_i) = 0$, then
\begin{align}\nonumber
&	\frac{1}{\sqrt{2\pi}} e^{-\frac{t_i^2}{2}} = \frac{e^{t_i}}{(1 + e^{t_i})^2}, \ i = 1, 2
	\Longleftrightarrow
	\frac{e^{-\frac{t_1^2}{2}}}{e^{-\frac{t_2^2}{2}}}
	= \frac{e^{t_1}}{e^{t_2}}\cdot \frac{(1 + e^{t_2})^2}{(1 + e^{t_1})^2}\\ \label{eq:u_prime}
\Longleftrightarrow &
	\frac{e^{\frac{(t_2+1)^2}{2}}}{e^{\frac{(t_1+1)^2}{2}}}
	= \left( \frac{1 + e^{t_2}}{1 + e^{t_1}} \right)^2
	\Longleftrightarrow
	\frac{(t_2 + 1)^2}{4} - \log(1 + e^{t_2}) = \frac{(t_1 + 1)^2}{4} - \log(1 + e^{t_1}).
\end{align}
Note that the derivative of the function $(t + 1)^2/4 - \log(1 + e^{t})$
is $(t+1)/2 - e^t/(1+e^t)$. It is straightforward to show that 
this derivative is strictly less than 0 for all $t<0$, so $(t + 1)^2/4 - \log(1 + e^{t})$
is a strictly decreasing function.
Thus \eqref{eq:u_prime} holds only if $t_1 = t_2$, which leads to a contradiction.

Hence for any $\epsilon > 0$, 
\begin{align*}
\int_{0}^{\infty} \beta_j p(\beta_j) g^{-1}_{\text{probit}}(- \epsilon \beta_j) d \beta_j
&	< \int_{0}^{\infty} \beta_j p(\beta_j) g^{-1}_{\text{logit}}(-
	\epsilon \beta_j) d \beta_j
	< \int_{0}^{\infty} \beta_j p(\beta_j) {e^{-\epsilon
      \beta_j}} d \beta_j \\
	=\int_{0}^{\infty} \frac{\beta_j e^{-\epsilon
      \beta_j}}{\pi \sigma_j (1 +  \beta_j^2 / \sigma_j^2)}  d \beta_j 
&	< \int_{0}^{\infty} \frac{\beta_j e^{-\epsilon
      \beta_j} }{\pi \sigma_j} d \beta_j 
= \frac{1}{\pi \sigma_j \epsilon^2} < \infty.
\end{align*}

Since the probit link is symmetric, i.e., 
$1 - g^{-1}_{\text{probit}}(\epsilon \beta_j) = 1 - \Phi(\epsilon \beta_j)  
= \Phi(-\epsilon \beta_j) = g^{-1}_{\text{probit}}(-\epsilon \beta_j)$,
we also have $\int_{0}^{\infty} \beta_j p(\beta_j) \left[ 1 - g^{-1}_{\text{probit}}(\epsilon \beta_j)\right] d \beta_j
< \infty$.
\end{proof}

\subsection{Proof of Corollary \ref{corollary:nonzero_prior_mean}}\label{proof_corollary:nonzero_prior_mean1}

To prove Corollary \ref{corollary:nonzero_prior_mean}, we mainly use a similar strategy to
the proof of Theorem \ref{theorem:existence}.
To show the necessary condition, we can use all of Appendix \ref{proof_th1_necessary}
 without modification, for both logistic and probit regression models. 
To show the sufficient condition, 
we can follow the same proof outline as in Appendix \ref{proof_th1_sufficient}, with some
minimal modification as described in the following proof.

\begin{proof}

First, we denote the vector of prior location parameters 
by $\boldsymbol\mu = (\mu_1, \mu_2, \ldots, \mu_p)^T$.
If we shift the coefficients $\boldsymbol\beta$ by $\boldsymbol\mu$ units, then 
\[
\boldsymbol\gamma \stackrel{\text{def}}{=} \boldsymbol\beta - \boldsymbol\mu ~ \Longrightarrow~
\gamma_j \stackrel{\text{ind}}{\sim} \text{C}(0, \sigma_j), \quad j = 1, 2, \ldots, p,
\]
that is, the resulting parameters $\gamma_j$ have independent Cauchy priors with location parameters being zero. 
Since the original 
parameter $\beta_j = \gamma_j + \mu_j$ for each $j = 1, 2, \ldots, p$,
the existence of $E(\beta_j \mid \mathbf{Y})$ is equivalent to the existence 
of $E(\gamma_j \mid \mathbf{Y})$. So we just need to show that
if $\mathbf{X}_j$ is not a solitary separator, then $E(\gamma_j \mid \mathbf{Y})$ exists.
For simplicity, here we just show that the positive term
\[
E(\gamma_j \mid \mathbf{y})^+ \stackrel{\text{def}}{=}
 \int_{0}^{\infty} \gamma_j p(\gamma_j) p(\mathbf{y} \mid \gamma_j) d \gamma_j
\]
has a finite upper bound. The other half of the result that the negative term
$E(\gamma_j \mid \mathbf{y})^-$ has a finite lower bound will follow with a similar derivation.

As in Appendix \ref{proof_th1_sufficient}, we first
assume that condition \ref{condition_a} is true, and define $\epsilon$ in the same way. 
For any $\gamma_j > 0$, we define a subset of the domain of $\bm{\gamma}_{(-j)} $
as follows
\[
G(\gamma_j) \stackrel{\text{def}}{=} \left\{  
\bm{\gamma}_{(-j)}  \in \mathbb{R}^{p-1}: \mathbf{x}_{i', (-j)}^T \bm{\gamma}_{(-j)} 
< \epsilon \gamma_j \right\},
\]
then for any $\bm{\gamma}_{(-j)} \in G(\gamma_j)$, 
$\mathbf{x}_{i'}^T \bm{\gamma} < - \epsilon \gamma_j$.
Since $f_1(\cdot)$ is an increasing function,  
\[
f_1(\mathbf{x}_{i'}^T \bm{\beta} ) = f_1(\mathbf{x}_{i'}^T \bm{\gamma} + \mathbf{x}_{i'}^T \bm{\mu}) 
< f_1(- \epsilon \gamma_j + \mathbf{x}_{i'}^T \bm{\mu}), \quad \text{for all } \bm{\gamma}_{(-j)} \in G(\gamma_j).
\]
A similar derivation to \eqref{eqn:pm_gamma1_plus} gives
\begin{align*} 
&	E(\gamma_j \mid \mathbf{y})^+\\ 
<~&	\int_{0}^{\infty} \gamma_j p(\gamma_j) 
	\left[ \int_{G(\gamma_j)} f_1(- \epsilon \gamma_j + \mathbf{x}_{i'}^T \bm{\mu}) 
	p(\bm{\gamma}_{(-j)}) d\bm{\gamma}_{(-j)} 
	+ \int_{\mathbb{R}^{p-1}\backslash G(\gamma_j)}  
	p(\bm{\gamma}_{(-j)}) d\bm{\gamma}_{(-j)} \right] d \gamma_j,
\end{align*}
where by \eqref{eqn:pm_gamma1_plus3} the first integral inside the bracket 
has an upper bound
\begin{equation}\label{eq:f1_mu}
\int_{G(\gamma_j)} f_1(- \epsilon \gamma_j + \mathbf{x}_{i'}^T \bm{\mu}) 
	p(\bm{\gamma}_{(-j)}) d\bm{\gamma}_{(-j)} 
< f_1(- \epsilon \gamma_j + \mathbf{x}_{i'}^T \bm{\mu}),
\end{equation}
and by \eqref{eqn:upperbound2} the second integral inside the bracket also has an upper bound
\[
\int_{\mathbb{R}^{p-1}\backslash G(\gamma_j)}  
	p(\bm{\gamma}_{(-j)}) d\bm{\gamma}_{(-j)}
<	\frac{|\mathbf{x}_{i', (-j)}|^T
  \bm{\sigma}_{(-j)}}{\pi\epsilon\gamma_j}.
\]

In logistic regression, the right hand side of \eqref{eq:f1_mu}
is further bounded 
\[
f_1(- \epsilon \gamma_j + \mathbf{x}_{i'}^T \bm{\mu})
= \frac{e^{- \epsilon \gamma_j + \mathbf{x}_{i'}^T \bm{\mu}}}
{1 + e^{- \epsilon \gamma_j + \mathbf{x}_{i'}^T \bm{\mu}}}
< e^{- \epsilon \gamma_j + \mathbf{x}_{i'}^T \bm{\mu}},
\]
and hence by \eqref{eqn:upperbound+}, 
\[
E(\gamma_j \mid \mathbf{y})^+
< \frac{e^{\mathbf{x}_{i'}^T \bm{\mu}}}{\pi \sigma_j \epsilon^2} + \frac{|\mathbf{x}_{i',
    (-j)}|^T \bm{\sigma}_{(-j)}}{2\pi\epsilon}.
\]

In 
probit regression, the function $f_1(\cdot)$ in the above derivations 
equals the standard normal cdf  $\Phi(\cdot)$. 
By \eqref{eq:probit_logit}, for any $\gamma_j > \mathbf{x}_{i'}^T \bm{\mu}/ \epsilon$, 
we have 
\[
\Phi \left(- \epsilon \gamma_j + \mathbf{x}_{i'}^T \bm{\mu}\right) 
< \frac{e^{- \epsilon \gamma_j + \mathbf{x}_{i'}^T \bm{\mu}}}{1 + e^{- \epsilon \gamma_j + \mathbf{x}_{i'}^T \bm{\mu}}}
< e^{- \epsilon \gamma_j + \mathbf{x}_{i'}^T \bm{\mu}}.
\]
Hence for $\mathbf{x}_{i'}^T \bm{\mu}/\epsilon > 0$ we have an upper bound
\begin{align*} 
&	E(\gamma_j \mid \mathbf{y})^+
	< \int_{0}^{\infty} \gamma_j p(\gamma_j) 
	\left[ \Phi \left(- \epsilon \gamma_j + \mathbf{x}_{i'}^T \bm{\mu}\right) 
	+ \frac{|\mathbf{x}_{i', (-j)}|^T
          \bm{\sigma}_{(-j)}}{\pi\epsilon\gamma_j} \right] d \gamma_j \\
=&	 \int_{0}^{\infty} \gamma_j p(\gamma_j) 
		\Phi \left(- \epsilon \gamma_j + \mathbf{x}_{i'}^T \bm{\mu}\right) d \gamma_j 
		+ \frac{|\mathbf{x}_{i',(-j)}|^T \bm{\sigma}_{(-j)}}{2\pi\epsilon}\\
=&	 \int_{0}^{ \mathbf{x}_{i'}^T \bm{\mu}/\epsilon} \gamma_j p(\gamma_j) 
		\Phi \left(- \epsilon \gamma_j + \mathbf{x}_{i'}^T \bm{\mu}\right) d \gamma_j 
		+ \int_{ \mathbf{x}_{i'}^T \bm{\mu}/ \epsilon}^{\infty} \gamma_j p(\gamma_j) 
		\Phi \left(- \epsilon \gamma_j + \mathbf{x}_{i'}^T \bm{\mu}\right) d \gamma_j 
		+ \frac{|\mathbf{x}_{i',(-j)}|^T \bm{\sigma}_{(-j)}}{2\pi\epsilon}\\
<&	 \int_{0}^{ \mathbf{x}_{i'}^T \bm{\mu}/\epsilon} \gamma_j p(\gamma_j) d \gamma_j 
		+ \int_{ \mathbf{x}_{i'}^T \bm{\mu}/ \epsilon}^{\infty} \gamma_j p(\gamma_j) 
		 e^{- \epsilon \gamma_j + \mathbf{x}_{i'}^T \bm{\mu}} d \gamma_j 
		+ \frac{|\mathbf{x}_{i',(-j)}|^T \bm{\sigma}_{(-j)}}{2\pi\epsilon}\\
<&	 \int_{0}^{ \mathbf{x}_{i'}^T \bm{\mu}/\epsilon} 
		\frac{\gamma_j}{\pi \sigma_j (1 +  \gamma_j^2 / \sigma_j^2)}  d \gamma_j 
		+ e^{\mathbf{x}_{i'}^T \bm{\mu}}\int_{ 0}^{\infty} 
		\frac{\gamma_j e^{- \epsilon \gamma_j }}{\pi \sigma_j (1 +  \gamma_j^2 / \sigma_j^2)} 
		  d \gamma_j 
		+ \frac{|\mathbf{x}_{i',(-j)}|^T \bm{\sigma}_{(-j)}}{2\pi\epsilon}\\
<&~	 \frac{\sigma_j}{2\pi} \log\left[1 +  \left(\frac{\mathbf{x}_{i'}^T \bm{\mu}}{\epsilon\sigma_j}  \right)^2 \right]
		+ e^{\mathbf{x}_{i'}^T \bm{\mu}}\int_{ 0}^{\infty} 
		\frac{\gamma_j e^{- \epsilon \gamma_j }}{\pi \sigma_j } 
		  d \gamma_j 
		+ \frac{|\mathbf{x}_{i',(-j)}|^T \bm{\sigma}_{(-j)}}{2\pi\epsilon}\\
=&~	 \frac{\sigma_j}{2\pi} \log\left[1 +  \left(\frac{\mathbf{x}_{i'}^T \bm{\mu}}{\epsilon\sigma_j}  \right)^2 \right]
		+ \frac{e^{\mathbf{x}_{i'}^T \bm{\mu}}}{\pi \sigma_j \epsilon^2}
		+ \frac{|\mathbf{x}_{i',(-j)}|^T \bm{\sigma}_{(-j)}}{2\pi\epsilon}.
\end{align*}
Note that a similar derivation also holds if $\mathbf{x}_{i'}^T \bm{\mu}/\epsilon < 0$.

On the other hand, if condition \ref{condition_b} 
is true, we can follow the same modification in Appendix \ref{proof_th1_sufficient}
to find upper bounds in a similar way.
\end{proof}

To show Theorem \ref{theorem:multivariate_Cauchy}, we decompose its proof 
into two parts: in Appendix \ref{proof_theorem:multivariate_Cauchy_no_separation} we show that 
all posterior means exist if there is no separation; then in Appendix 
\ref{proof_theorem:multivariate_Cauchy_complete_separation} we show that 
under a multivariate Cauchy prior, 
none of the posterior means exist if there is complete separation.

\subsection{Proof of Theorem \ref{theorem:multivariate_Cauchy}, in the case of no separation}
\label{proof_theorem:multivariate_Cauchy_no_separation}
\begin{proof}
For any $j = 1, 2, \ldots, p$, to show that $E(\beta_j \mid \mathbf{y})$ exists, 
we just need to show the positive term $E(\beta_j \mid \mathbf{y})^+$ 
in \eqref{eq:posterior_mean_beta_j} has an upper bound,
because the negative term $E(\beta_j \mid \mathbf{y})^-$ in \eqref{eq:posterior_mean_beta_j}
having a lower bound follows a similar derivation.

When working on $E(\beta_j \mid \mathbf{y})^+$, we only need to consider positive $\beta_j$.
Denote a new $p-1$ dimensional variable $\bm{\gamma} = \bm{\beta}_{(-j)}/ \beta_j$, 
then for $i = 1, 2, \ldots, n$,
\[
\mathbf{x}_i^T \bm{\beta} = \beta_j\left( x_{i,j} + \mathbf{x}_{i,(-j)}^T \bm{\gamma} \right).
\]
If there is no separation, for any $\bm{\gamma} \in \mathbb{R}^{p-1}$,
there exists at least one $i \in \{1, 2, \ldots, n\}$, such that
\begin{equation}\label{eq:no_separtion_1}
 x_{i,j} + \mathbf{x}_{i,(-j)}^T \bm{\gamma}  < 0, \text{ if } i \in A_1, \text{ or }
 x_{i,j} + \mathbf{x}_{i,(-j)}^T \bm{\gamma}  > 0, \text{ if } i \in A_0.
\end{equation}
For each $i = 1, 2, \ldots, n$, denote the vector $\mathbf{z}_i$ and the function $h_i(\cdot)$ as follows,
\begin{equation}\label{eq:z_h}
\mathbf{z}_i \stackrel{\text{def}}{=} \begin{cases}
	\mathbf{x}_i  & \text{ if } i  \in A_1 \\
	-\mathbf{x}_i  & \text{ if } i  \in A_0 \\
	\end{cases}, \quad
h_i(\bm{\gamma}) 	\stackrel{\text{def}}{=} z_{i, j} + \mathbf{z}_{i, (-j)}^T \bm{\gamma},
\end{equation}
then \eqref{eq:no_separtion_1} can be rewritten as
$h_i (\bm{\gamma}) < 0$.
Denote for $i = 1, 2, \ldots, n$,
\[
B_i \stackrel{\text{def}}{=}  \{\bm{\gamma}: h_i(\bm{\gamma}) < 0\}.
\]
Then each $B_i$ is a non-empty subset
of $\mathbb{R}^{p-1}$, unless $z_{i,j} \geq 0$ and $\mathbf{z}_{i, (-j)} = \mathbf{0}$.
Let $\mathcal{I} = \{i: B_i \neq \o\}$ denote the set of indices $i$, for which the corresponding $B_i$ are non-empty.
Because there is no separation,
\begin{equation}\label{eq:B_i_union}
\bigcup_{i \in \mathcal{I}} B_i = \mathbb{R}^{p-1}.
\end{equation}
Hence, the set $\mathcal{I}$ is non-empty.
We denote its size by $q \stackrel{\text{def}}{=} |\mathcal{I}|$, and rewrite 
$\mathcal{I} = \{i_1, i_2, \ldots, i_q\}$.

Now we show that there exist positive constants $\epsilon_{i_1}, \epsilon_{i_2}, \ldots, \epsilon_{i_q}$,
such that 
\begin{equation}\label{eq:tildeB_i2}
\bigcup_{k=1}^q \tilde{B}_{i_k} = \mathbb{R}^{p-1},  
\end{equation}
where
\begin{equation}\label{eq:tildeB_i}
\tilde{B}_{i_k} \stackrel{\text{def}}{=} \{\bm{\gamma}: h_{i_k}(\bm{\gamma}) < -\epsilon_{i_k}\}, 
\end{equation}
are subsets of the corresponding $B_{i_k}$, for all $k = 1, 2, \ldots, q$.

If there exists an $i_k \in \mathcal{I}$ such that $z_{i_k,j} < 0$ and $\mathbf{z}_{i_k, (-j)} = \mathbf{0}$,
then $B_{i_k} = \mathbb{R}^{p-1}$. In this case, we just need to let $\epsilon_{i_k} = - z_{i_k,j} /2$, and 
$\epsilon_{i_r} = M$, for all $r \neq k$, where $M$ is an arbitrary positive number.
Under this choice of $\epsilon_{i}$'s, the sets $\tilde{B}_i$'s defined by \eqref{eq:tildeB_i} satisfy \eqref{eq:tildeB_i2}.

If, on the other hand, $\mathbf{z}_{i_k, (-j)} \neq \mathbf{0}$ for all $i_k \in \mathcal{I}$, i.e., 
all $B_{i_k}$ are open half spaces in $\mathbb{R}^{p-1}$, then
we can find the constants $\epsilon_{i_1}, \epsilon_{i_2}, \ldots, \epsilon_{i_q}$ sequentially.
For $i_1$, if $\bigcup_{k=2}^q B_{i_k} = \mathbb{R}^{p-1}$,  
we can set $\epsilon_{i_1} = M$. Then the resulting $\tilde{B}_{i_1}$ defined by \eqref{eq:tildeB_i} satisfies
\begin{equation}\label{eq:B1_modified}
\tilde{B}_{i_1} \cup B_{i_2} \cup B_{i_3} \cup \cdots \cup B_{i_q} = \mathbb{R}^{p-1}.
\end{equation}
If $\bigcup_{k=2}^q B_{i_k} \neq \mathbb{R}^{p-1}$, then \eqref{eq:B_i_union} suggests
\begin{equation}\label{eq:B1_supset}
B_{i_1} \supset \left(\bigcup_{k=2}^q B_{i_k}  \right)^c = \bigcap_{k=2}^q B_{i_k}^c.
\end{equation}
For \eqref{eq:B1_modified} to hold,  
we just need to find an positive $\epsilon_{i_1}$ such that the resulting $\tilde{B}_{i_1}$ has
 $\bigcap_{k=2}^q B_{i_k}^c$
as a subset, i.e., $-\epsilon_{i_1}$ should be larger than the maximum of $h_{i_1}(\bm{\gamma})$
over the polyhedral region $\bm{\gamma} \in \bigcap_{k=2}^q B_{i_k}^c$. 
Note that maximizing $h_{i_1}(\bm{\gamma})$ over the polyhedron
can be represented as a linear programming question,
\begin{align}\label{eq:lp1}
\text{maximize} \quad & z_{i_1, j} + \mathbf{z}_{i_1, (-j)}^T\bm{\gamma}\\ \nonumber
\text{subject to} \quad &  \mathbf{z}_{i_2, (-j)}^T\bm{\gamma} \geq - z_{i_2, j}\\  \nonumber
	& \vdots\\ \nonumber
	&  \mathbf{z}_{i_q, (-j)}^T\bm{\gamma} \geq - z_{i_q, j}.
\end{align}
By \citet[pp.\ 67, Corollary 2.3]{Bertsimas_Tsitsiklis_1997}, for any linear programming
problem over a non-empty polyhedron, including the one in \eqref{eq:lp1}
to maximize $h_{i_1}(\bm{\gamma}) = z_{i_1, j} + \mathbf{z}_{i_1, (-j)}^T\bm{\gamma}$, 
either the optimal $h_{i_1}(\bm{\gamma}) = \infty$,
or there exists an optimal solution, $\bm{\gamma}^*$. 
Here, the latter case always occurs, 
because by \eqref{eq:B1_supset}, the maximum of $h_{i_1}(\bm{\gamma})$
over the polyhedron $\bigcap_{k=2}^q B_{i_k}^c$ does not exceed zero,
so it does not go to infinity.
Hence, we just need to let 
\[
\epsilon_{i_1} = -\frac{1}{2}\left[ 
	\max_{\bm{\gamma} \in \bigcap_{k=2}^q B_{i_k}^c} z_{i_1, j} + \mathbf{z}_{i_1, (-j)}^T\bm{\gamma} \right]
	= -\frac{z_{i_1, j} + \mathbf{z}_{i_1, (-j)}^T\bm{\gamma}^*}{2},
\]
so that the resulting $\tilde{B}_{i_1}= \{\bm{\gamma}: h_{i_1}(\bm{\gamma}) < -\epsilon_{i_1}\}$
contains $\bigcap_{k=2}^q B_{i_k}^c$ as a subset, which yields \eqref{eq:B1_modified}.
After finding $\epsilon_{i_1}$, we can apply similar procedures sequentially, 
to find positive values $\epsilon_{i_k}$, for $k = 2, 3, \ldots,q$, such that 
\[
\tilde{B}_{i_1} \cup \cdots \cup \tilde{B}_{i_k} \cup B_{i_{k+1}} \cup \cdots \cup B_{i_q} = \mathbb{R}^{p-1}.
\]
After identifying all $\epsilon_{i_k}$'s, the resulting $\tilde{B}_{i_k}$'s satisfy \eqref{eq:tildeB_i2}.
Note that the choice of  $\epsilon_{i_k}$'s only depend on the data $\mathbf{z}_i$, $i = 1, 2, \ldots, n$, 
so they are constants given the observed data.

For each $k = 1, 2, \ldots, q$, 
next we show that for any $\bm{\gamma}\in \tilde{B}_{i_k}$, the likelihood function of the $i_k$th observation
is bounded above by $(\beta_j \epsilon_{i_k} e)^{-1}$.
This is because in a logistic regression, if $i_k \in A_1$, then 
\begin{equation}\label{eq:f1_upper_bound1}
p(y_{i_k} \mid \beta_j, \bm{\gamma}) = f_1\left(\beta_j h_{i_k}(\bm{\gamma})\right)
= \frac{e^{\beta_j h_{i_k}(\bm{\gamma})}}{1 + e^{\beta_j h_{i_k}(\bm{\gamma})}} < e^{\beta_j h_{i_k}(\bm{\gamma})}
< e^{-\beta_j \epsilon_{i_k}} \leq \frac{1}{\beta_j \epsilon_{i_k}e},
\end{equation}
if $i_k \in A_0$, then
\begin{equation}\label{eq:f1_upper_bound0}
p(y_{i_k} \mid \beta_j, \bm{\gamma}) = f_0\left(-\beta_j h_{i_k}(\bm{\gamma})\right)
= \frac{1}{1 + e^{-\beta_j h_{i_k}(\bm{\gamma})}} < e^{\beta_j h_{i_k}(\bm{\gamma})}
< e^{-\beta_j \epsilon_{i_k}} \leq \frac{1}{\beta_j \epsilon_{i_k}e}.
\end{equation}
Here, the last inequality holds because $e^{-t} \leq \frac{e^{-1}}{t}$ for any $t > 0$.
By \eqref{eq:probit_logit}, in a probit regression model, 
the inequalities \eqref{eq:f1_upper_bound1} and \eqref{eq:f1_upper_bound0} also hold.

Now we show that the positive term $E(\beta_j \mid \mathbf{y})^+$ has a finite upper bound.
\begin{align*}
E(\beta_j \mid \mathbf{y})^+
&	=  \int_{0}^{\infty} \beta_j \int_{\mathbb{R}^{p-1}} 
	p(\mathbf{y} \mid \beta_j, \bm{\gamma}) p(\beta_j, \bm{\gamma}) d \bm{\gamma} d \beta_j\\
& 	\leq  \int_{0}^{\infty} \beta_j \sum_{k=1}^q \int_{\tilde{B}_{i_k}} 
	p(\mathbf{y} \mid \beta_j, \bm{\gamma}) p(\beta_j, \bm{\gamma}) d \bm{\gamma} d \beta_j\\
& 	<  \int_{0}^{\infty} \beta_j \sum_{k=1}^q \int_{\tilde{B}_{i_k}} 
	p(y_{i_k} \mid \beta_j, \bm{\gamma}) p(\beta_j, \bm{\gamma}) d \bm{\gamma} d \beta_j\\
&	\leq \int_{0}^{\infty} \beta_j \sum_{k=1}^q \int_{\tilde{B}_{i_k}} 
	\frac{1}{\beta_j \epsilon_{i_k} e} p(\beta_j, \bm{\gamma}) d \bm{\gamma} d \beta_j\\
&	= \sum_{k=1}^q   \frac{1}{\epsilon_{i_k} e} \int_{0}^{\infty} \int_{\tilde{B}_{i_k}} 
	 p(\beta_j, \bm{\gamma}) d \bm{\gamma} d \beta_j\\	
&	\leq \sum_{k=1}^q   \frac{1}{\epsilon_{i_k} e} \int_{0}^{\infty} \int_{\mathbb{R}^{p-1}} 
	 p(\beta_j, \bm{\gamma}) d \bm{\gamma} d \beta_j
	 < \sum_{k=1}^q   \frac{1}{\epsilon_{i_k} e}.
\end{align*}
\end{proof}

Note that in Appendix \ref{proof_theorem:multivariate_Cauchy_no_separation},
the specific formula of the prior density of $\boldsymbol\beta$ is not used.
Therefore, if there is no separation in logistic or probit regression, 
posterior means of all coefficients exist under all proper prior distributions.

\subsection{Proof of Theorem \ref{theorem:multivariate_Cauchy}, in the case of complete separation} \label{proof_theorem:multivariate_Cauchy_complete_separation}
\begin{proof}
Here we show that if there is complete separation, then none of the posterior means 
$E(\beta_j \mid \mathbf{y})$ exist, for $j = 1, 2, \ldots, p$.
Using the notation $\mathbf{z}_i$, defined in \eqref{eq:z_h}, 
we rewrite the set of all vectors 
satisfying
the complete separation condition \eqref{eqn:compsep} as
\[
\mathcal{C} = \bigcap_{i=1}^n \left\{\bm{\beta}\in \mathbb{R}^{p}: \mathbf{z}_i^T \bm{\beta} > 0\right\}.
\]
According to \citet{Albert_Anderson_1984}, $\mathcal{C}$ is a convex cone; moreover,
if $\bm{\beta} \in \mathcal{C}$, then
$\bm{\beta}  + \bm{\delta} \in \mathcal{C}$ for any $\bm{\delta} \in \mathbb{R}^{p}$ that is small enough.
Hence, the open set $\mathcal{C}$, as a subset of the $\mathbb{R}^p$ Euclidean space, 
has positive Lebesgue measure. 

To show that $E(\beta_j \mid \mathbf{y})$ does not exist, if $\mathcal{C}$ projects
on the positive half of the $\beta_j$ axis, we will show that $E(\beta_j \mid \mathbf{y})^+$
diverges, otherwise, we will show that $E(\beta_j \mid \mathbf{y})^-$ diverges
(if both, then showing either is sufficient).
Now we assume that the former is true, and denote the intersection 
\[
\mathcal{C}_j^+ \stackrel{\text{def}}{=} \mathcal{C} \cap \{\bm{\beta}\in \mathbb{R}^{p}: \beta_j > 0\},
\]
which is again an open convex cone. Since $\mathcal{C}_j^+$ has positive measure in $\mathbb{R}^p$,
under the change of variable from $(\beta_j, \bm{\beta}_{(-j)})$ to $(\beta_j, \bm{\gamma})$, where
$\bm{\beta}_{(-j)} = \beta_j\bm{\gamma}$, there exists an open set 
$\tilde{\mathcal{C}}_j^+ \in \mathbb{R}^{p-1}$ such that  $\mathcal{C}_j^+$ can be written as
\[
\mathcal{C}_j^+ = \{(\beta_j, \beta_j \bm{\gamma}): \beta_j > 0, \bm{\gamma}\in  \tilde{\mathcal{C}}_j^+  \}.
\]
Suppose that $\bm{\gamma}$ can be written as 
$(\gamma_1, \ldots, \gamma_{j-1}, \gamma_{j+1}, \ldots, \gamma_p)^T$.
We define a variant of it by $\tilde{\bm{\gamma}} \stackrel{\text{def}}{=}
(\gamma_1, \ldots, \gamma_{j-1}, 1, \gamma_{j+1}, \ldots, \gamma_p)^T$,
such that $\bm{\beta} = \beta_j \tilde{\bm{\gamma}}$.
Under the multivariate Cauchy prior \eqref{eq:multi_Cauchy_prior},
the induced prior distribution of $(\beta_j, \bm{\gamma})$ is
\begin{align*}
p(\beta_j, \bm{\gamma}) 
&	\propto \frac{\beta_j^{p-1}}
	{\left[1 + \left(\beta_j \tilde{\bm{\gamma}} - \bm{\mu}\right)^T 
	\boldsymbol\Sigma^{-1} 
	\left(\beta_j \tilde{\bm{\gamma}} - \bm{\mu}\right)\right]^{\frac{p+1}{2}}}\\
&	= \frac{\beta_j^{p-1}}
	{\left[\left(\tilde{\bm{\gamma}}^T \boldsymbol\Sigma^{-1} \tilde{\bm{\gamma}}\right) \beta_j^2
	- 2 \left(\tilde{\bm{\gamma}}^T \boldsymbol\Sigma^{-1} \bm{\mu}\right) \beta_j
	+ \left(\bm{\mu}^T \boldsymbol\Sigma^{-1} \bm{\mu} + 1\right)\right]^{\frac{p+1}{2}}}.
\end{align*}

Inside $\tilde{\mathcal{C}}_j^+$, there must exist a closed rectangular box,
denoted by $\tilde{\mathcal{D}}_j^+ = \{\bm{\gamma}: \gamma_k \in [l_k, u_k], k = 1, \ldots, j-1, j+ 1, \ldots, p\} 
\subset \tilde{\mathcal{C}}_j^+$. By \citet[pp.\ 142, Corollary 6.57]{Browder_1996}, 
a continuous function takes its maximum and minimum on a compact set.
Since $\mathcal{D}_j^+$ is a compact set (closed and bounded
in $\mathbb{R}^{p-1}$), 
\begin{equation}
a \stackrel{\text{def}}{=} \max_{\bm{\gamma} \in \tilde{\mathcal{D}}_j^+} 
\tilde{\bm{\gamma}}^T \boldsymbol\Sigma^{-1} \tilde{\bm{\gamma}}, \quad
b \stackrel{\text{def}}{=} \min_{\bm{\gamma} \in \tilde{\mathcal{D}}_j^+} 
\tilde{\bm{\gamma}}^T \boldsymbol\Sigma^{-1} \bm{\mu}
\label{eq:def_a_b}
\end{equation}
both exist.

Recall that for all $\bm{\gamma}\in \tilde{\mathcal{C}}_j^+$
(hence including all elements in $\mathcal{D}_j^+$), 
$z_{i,j} + \mathbf{z}_{i, (-j)}^T \bm{\gamma} >0$; equivalently, 
if $i \in A_1$, then $x_{i,j} + \mathbf{x}_{i,(-j)}^T \bm{\gamma} > 0$, 
and if $i \in A_0$, then $x_{i,j} + \mathbf{x}_{i,(-j)}^T \bm{\gamma} < 0$. 
Now we show that in both logistic and probit regressions,
the positive term $E(\beta_j \mid \mathbf{y})^+$ diverges.
\begin{align}\nonumber
&	E(\beta_j \mid \mathbf{y})^+
	=  \int_{0}^{\infty} \beta_j \int_{\mathbb{R}^{p-1}} p(\beta_j, \bm{\gamma}) 
	p(\mathbf{y} \mid \beta_j, \bm{\gamma}) d \bm{\gamma} d \beta_j\\ \nonumber
\geq& 	  \int_{0}^{\infty} \beta_j  \int_{\tilde{\mathcal{D}}_j^+} p(\beta_j, \bm{\gamma})  
	\prod_{i \in A_1}f_1(\beta_j (x_{i,j} + \mathbf{x}_{i,(-j)}^T \bm{\gamma})) 
	\prod_{k \in A_0}f_0(\beta_j (x_{k,j} + \mathbf{x}_{k,(-j)}^T \bm{\gamma}))
	d \bm{\gamma} d \beta_j\\ \nonumber
\geq& 	  \int_{0}^{\infty} \beta_j  \int_{\tilde{\mathcal{D}}_j^+} p(\beta_j, \bm{\gamma})  
	\prod_{i \in A_1}f_1(0) 
	\prod_{k \in A_0}f_0(0)
	d \bm{\gamma} d \beta_j\\ \nonumber
=& 	 \left(\frac{1}{2}\right)^n  \int_{0}^{\infty} \beta_j  
	\int_{\tilde{\mathcal{D}}_j^+} p(\beta_j, \bm{\gamma})  d \bm{\gamma} d \beta_j\\ \nonumber
=&	~C \int_{0}^{\infty} \beta_j  
	\int_{\tilde{\mathcal{D}}_j^+} \frac{\beta_j^{p-1}}
	{\left[\left(\tilde{\bm{\gamma}}^T \boldsymbol\Sigma^{-1} \tilde{\bm{\gamma}}\right) \beta_j^2
	- 2 \left(\tilde{\bm{\gamma}}^T \boldsymbol\Sigma^{-1} \bm{\mu}\right) \beta_j
	+ \left(\bm{\mu}^T \boldsymbol\Sigma^{-1} \bm{\mu} + 1\right)\right]^{\frac{p+1}{2}}}  
	d \bm{\gamma} d \beta_j\\ \nonumber
\geq&	~C  \int_{0}^{\infty}  
	\int_{\tilde{\mathcal{D}}_j^+} \frac{\beta_j^{p}}
	{\left[a\beta_j^2 - 2b \beta_j + \left(\bm{\mu}^T \boldsymbol\Sigma^{-1} \bm{\mu} + 1\right) \right]^{\frac{p+1}{2}}}  
	d \bm{\gamma} d \beta_j\\ \label{eq:int_diverge}
=&	~C \left[\prod_{k=1}^{p-1}(u_k - l_k)\right]\int_{0}^{\infty}  
	 \frac{\beta_j^{p}}
	{\left[a(\beta_j - b/a)^2 + c\right]^{\frac{p+1}{2}}}  
	d \beta_j \\ \nonumber
=& ~ \infty,
\end{align}
where $C$ is a positive constant, $c$ is a constant, and $a$ and $b$ have been defined previously in \eqref{eq:def_a_b}.

On the other hand, if the set $\mathcal{C}$ of complete separation vectors
only projects on the negative half of the $\beta_j$ axis, 
following a similar deviation, 
we can show that $E(\beta_j \mid \mathbf{y})^-$ diverges to $-\infty$.

\end{proof}

\bibliography{Cauchy.bib}

\begin{thebibliography}{45}
\newcommand{\enquote}[1]{``#1''}
\expandafter\ifx\csname natexlab\endcsname\relax\def\natexlab#1{#1}\fi
\expandafter\ifx\csname url\endcsname\relax
  \def\url#1{{\tt #1}}\fi
\expandafter\ifx\csname urlprefix\endcsname\relax\def\urlprefix{URL }\fi
\ifx\endbibitem\undefined \let\endbibitem\relax\fi

\bibitem[{Albert and Anderson(1984)}]{Albert_Anderson_1984}
Albert, A. and Anderson, J.~A. (1984).
\newblock \enquote{On the Existence of Maximum Likelihood Estimates in Logistic
  Regression Models.}
\newblock {\em Biometrika\/}, 71(1): 1--10.
\endbibitem

\bibitem[{Bardenet et~al.(2014)Bardenet, Doucet, and
  Holmes}]{Bardenet_etal_2014}
Bardenet, R., Doucet, A., and Holmes, C. (2014).
\newblock \enquote{Towards Scaling up {M}arkov Chain {M}onte {C}arlo: An
  Adaptive Subsampling Approach.}
\newblock {\em Proceedings of the 31st International Conference on Machine
  Learning (ICML-14)\/}, 405--413.
\endbibitem

\bibitem[{Bertsimas and Tsitsiklis(1997)}]{Bertsimas_Tsitsiklis_1997}
Bertsimas, D. and Tsitsiklis, J.~N. (1997).
\newblock {\em Introduction to Linear Optimization\/}.
\newblock Athena Scientific Belmont, MA.
\endbibitem

\bibitem[{Bickel and Doksum(2001)}]{Bickel_Doksum_2001}
Bickel, P.~J. and Doksum, K.~A. (2001).
\newblock {\em Mathematical Statistics, volume I\/}.
\newblock Prentice Hall Englewood Cliffs, NJ.
\endbibitem

\bibitem[{Blattenberger and Lad(1985)}]{Blattenberger_Lad_1985}
Blattenberger, G. and Lad, F. (1985).
\newblock \enquote{Separating the Brier Score into Calibration and Refinement
  Components: A Graphical Exposition.}
\newblock {\em The American Statistician\/}, 39(1): 26--32.
\endbibitem

\bibitem[{Brier(1950)}]{Brier_1950}
Brier, G.~W. (1950).
\newblock \enquote{Verification of Forecasts Expressed in Terms of
  Probability.}
\newblock {\em Monthly Weather Review\/}, 78: 1--3.
\endbibitem

\bibitem[{Browder(1996)}]{Browder_1996}
Browder, A. (1996).
\newblock {\em Mathematical Analysis: An Introduction\/}.
\newblock Springer-Verlag.
\endbibitem

\bibitem[{Carpenter et~al.(2016)Carpenter, Gelman, Hoffman, Lee, Goodrich,
  Betancourt, Brubaker, Guo, Li, and
  Riddell}]{Carp:Gelm:Hoff:Lee:Good:Beta:Brub:Guo:Li:Ridd:2016}
Carpenter, B., Gelman, A., Hoffman, M., Lee, D., Goodrich, B., Betancourt, M.,
  Brubaker, A., Michael, Guo, J., Li, P., and Riddell, A. (2016).
\newblock \enquote{Stan: A Probabilistic Programming Language.}
\newblock {\em Journal of Statistical Software\/}, in press.
\endbibitem

\bibitem[{Casella and Berger(1990)}]{Casella_Berger_1990}
Casella, G. and Berger, R.~L. (1990).
\newblock {\em Statistical Inference\/}.
\newblock Duxbury Press.
\endbibitem

\bibitem[{Chen and Shao(2001)}]{Chen_Shao_2001}
Chen, M.-H. and Shao, Q.-M. (2001).
\newblock \enquote{Propriety of Posterior Distribution for Dichotomous Quantal
  Response Models.}
\newblock {\em Proceedings of the American Mathematical Society\/}, 129(1):
  293--302.
\endbibitem

\bibitem[{Choi and Hobert(2013)}]{Choi_Hobert_2013}
Choi, H.~M. and Hobert, J.~P. (2013).
\newblock \enquote{The Polya-Gamma Gibbs Sampler for {B}ayesian Logistic
  Regression is Uniformly Ergodic.}
\newblock {\em Electronic Journal of Statistics\/}, 7(2054-2064).
\endbibitem

\bibitem[{Chopin and Ridgway(2015)}]{Chopin_Ridgway_2015}
Chopin, N. and Ridgway, J. (2015).
\newblock \enquote{Leave Pima Indians Alone: Binary Regression as a Benchmark
  for {B}ayesian Computation.}
\newblock {\em arxiv.org\/}.
\endbibitem

\bibitem[{Clogg et~al.(1991)Clogg, Rubin, Schenker, Schultz, and
  Weidman}]{Clogg_etal_1991}
Clogg, C.~C., Rubin, D.~B., Schenker, N., Schultz, B., and Weidman, L. (1991).
\newblock \enquote{Multiple Imputation of Industry and Occupation Codes in
  Census Public-Use Samples Using {B}ayesian Logistic Regression.}
\newblock {\em Journal of the American Statistical Association\/}, 86(413):
  68--78.
\endbibitem

\bibitem[{Dawid(1973)}]{Dawid_1973}
Dawid, A.~P. (1973).
\newblock \enquote{Posterior Expectations for Large Observations.}
\newblock {\em Biometrika\/}, 60: 664--666.
\endbibitem

\bibitem[{Fern{\'a}ndez and Steel(2000)}]{Fernandez_Steel_2000}
Fern{\'a}ndez, C. and Steel, M.~F. (2000).
\newblock \enquote{Bayesian Regression Analysis with Scale Mixtures of
  Normals.}
\newblock {\em Econometric Theory\/}, 16(80-101).
\endbibitem

\bibitem[{Firth(1993)}]{Firth_1993}
Firth, D. (1993).
\newblock \enquote{Bias Reduction of Maximum Likelihood Estimates.}
\newblock {\em Biometrika\/}, 80(1): 27--38.
\endbibitem

\bibitem[{Fouskakis et~al.(2009)Fouskakis, Ntzoufras, and
  Draper}]{Fouskakis_etal_2009}
Fouskakis, D., Ntzoufras, I., and Draper, D. (2009).
\newblock \enquote{Bayesian Variable Selection Using Cost-Adjusted BIC, with
  Application to Cost-Effective Measurement of Quality of Health Care.}
\newblock {\em The Annals of Applied Statistics\/}, 3(2): 663--690.
\endbibitem

\bibitem[{Gelman et~al.(2008)Gelman, Jakulin, Pittau, and
  Su}]{Gelman_etal_2008}
Gelman, A., Jakulin, A., Pittau, M., and Su, Y. (2008).
\newblock \enquote{A Weakly Informative Default Prior Distribution for Logistic
  and Other Regression Models.}
\newblock {\em The Annals of Applied Statistics\/}, 2(4): 1360--1383.
\endbibitem

\bibitem[{Gelman et~al.(2015)Gelman, Su, Yajima, Hill, Pittau, Kerman, Zheng,
  and Dorie}]{Gelman_etal_2015}
Gelman, A., Su, Y.-S., Yajima, M., Hill, J., Pittau, M.~G., Kerman, J., Zheng,
  T., and Dorie, V. (2015).
\newblock {\em {arm: Data Analysis Using Regression and Multilevel/Hierarchical
  Models}\/}.
\newblock {R} package version 1.8-5.
\newline\urlprefix\url{http://CRAN.R-project.org/package=arm}
\endbibitem

\bibitem[{Ghosh and Clyde(2011)}]{Ghos:Clyd:2011}
Ghosh, J. and Clyde, M.~A. (2011).
\newblock \enquote{Rao-{B}lackwellization for {B}ayesian Variable Selection and
  Model Averaging in Linear and Binary Regression: A Novel Data Augmentation
  Approach.}
\newblock {\em Journal of the American Statistical Association\/}, 106(495):
  1041--1052.
\endbibitem

\bibitem[{Ghosh et~al.(2011)Ghosh, Herring, and
  Siega-Riz}]{Ghos:Herr:Sieg:2011}
Ghosh, J., Herring, A.~H., and Siega-Riz, A.~M. (2011).
\newblock \enquote{Bayesian Variable Selection for Latent Class Models.}
\newblock {\em Biometrics\/}, 67: 917--925.
\endbibitem

\bibitem[{Ghosh and Reiter(2013)}]{Ghos:Reit:2013}
Ghosh, J. and Reiter, J.~P. (2013).
\newblock \enquote{{Secure {B}ayesian Model Averaging for Horizontally
  Partitioned Data}.}
\newblock {\em Statistics and Computing\/}, 23: 311--322.
\endbibitem

\bibitem[{Gramacy and Polson(2012)}]{Gramacy_Polson_2012}
Gramacy, R.~B. and Polson, N.~G. (2012).
\newblock \enquote{Simulation-Based Regularized Logistic Regression.}
\newblock {\em Bayesian Analysis\/}, 7(3): 567--590.
\endbibitem

\bibitem[{Hanson et~al.(2014)Hanson, Branscum, and Johnson}]{Hanson_etal_2014}
Hanson, T.~E., Branscum, A.~J., and Johnson, W.~O. (2014).
\newblock \enquote{Informative g-Priors for Logistic Regression.}
\newblock {\em Bayesian Analysis\/}, 9(3): 597--612.
\endbibitem

\bibitem[{Heinze(2006)}]{Heinze_2006}
Heinze, G. (2006).
\newblock \enquote{A Comparative Investigation of Methods for Logistic
  Regression with Separated or Nearly Separated Data.}
\newblock {\em Statistics in Medicine\/}, 25: 4216--4226.
\endbibitem

\bibitem[{Heinze and Schemper(2002)}]{Heinze_Schemper_2002}
Heinze, G. and Schemper, M. (2002).
\newblock \enquote{A Solution to the Problem of Separation in Logistic
  Regression.}
\newblock {\em Statistics in Medicine\/}, 21: 2409--2419.
\endbibitem

\bibitem[{Hoffman and Gelman(2014)}]{Hoff:Gelm:2014}
Hoffman, M.~D. and Gelman, A. (2014).
\newblock \enquote{The No-U-Turn Sampler: Adaptively Setting Path Lengths in
  {H}amiltonian {M}onte {C}arlo.}
\newblock {\em The Journal of Machine Learning Research\/}, 15(1): 1593--1623.
\endbibitem

\bibitem[{Holmes and Held(2006)}]{Holmes_Held_2006}
Holmes, C.~C. and Held, L. (2006).
\newblock \enquote{Bayesian Auxiliary Variable Models for Binary and
  Multinomial Regression.}
\newblock {\em Bayesian Analysis\/}, 1(1): 145--168.
\endbibitem

\bibitem[{Ibrahim and Laud(1991)}]{Ibrahim_Laud_1991}
Ibrahim, J.~G. and Laud, P.~W. (1991).
\newblock \enquote{On {B}ayesian Analysis of Generalized Linear Models using
  {J}effreys's Prior.}
\newblock {\em Journal of the American Statistical Association\/}, 86(416):
  981--986.
\endbibitem

\bibitem[{Jeffreys(1961)}]{Jeffreys_1961}
Jeffreys, H. (1961).
\newblock {\em Theory of Probability\/}.
\newblock Oxford Univ. Press.
\endbibitem

\bibitem[{Kurgan et~al.(2001)Kurgan, Cios, Tadeusiewicz, Ogiela, and
  Goodenday}]{Kurg:Cios:Tade:Ogie:Good:2001}
Kurgan, L., Cios, K., Tadeusiewicz, R., Ogiela, M., and Goodenday, L. (2001).
\newblock \enquote{{Knowledge Discovery Approach to Automated Cardiac SPECT
  Diagnosis}.}
\newblock {\em Artificial Intelligence in Medicine\/}, 23:2: 149--169.
\endbibitem

\bibitem[{Li and Clyde(2015)}]{Li:Clyd:2015}
Li, Y. and Clyde, M.~A. (2015).
\newblock \enquote{Mixtures of $g$-Priors in Generalized Linear Models.}
\newblock {\em arxiv.org\/}.
\endbibitem

\bibitem[{Liu(2004)}]{Liu_2004}
Liu, C. (2004).
\newblock \enquote{Robit Regression: A Simple Robust Alternative to Logistic
  and Probit Regression.}
\newblock In Gelman, A. and Meng, X. (eds.), {\em Applied {B}ayesian Modeling
  and Casual Inference from Incomplete-Data Perspectives\/}, 227--238. Wiley,
  London.
\endbibitem

\bibitem[{McCullagh and Nelder(1989)}]{McCullagh_Nelder_1989}
McCullagh, P. and Nelder, J. (1989).
\newblock {\em Generalized Linear Models\/}.
\newblock Chapman and Hall.
\endbibitem

\bibitem[{Mitra and Dunson(2010)}]{Mitr:Duns:2010}
Mitra, R. and Dunson, D.~B. (2010).
\newblock \enquote{Two Level Stochastic Search Variable Selection in {GLM}s
  with Missing Predictors.}
\newblock {\em International Journal of Biostatistics\/}, 6(1): Article 33.
\endbibitem

\bibitem[{Neal(2003)}]{Neal:2003}
Neal, R.~M. (2003).
\newblock \enquote{Slice Samlping.}
\newblock {\em The Annals of Statistics\/}, 31(3): 705--767.
\endbibitem

\bibitem[{Neal(2011)}]{Neal:2011}
--- (2011).
\newblock \enquote{{MCMC} using {H}amiltonian Dynamics.}
\newblock In Brooks, S., Gelman, A., Jones, G., and Meng, X.-L. (eds.), {\em
  Handbook of Markov Chain Monte Carlo\/}. Chapman \& Hall / CRC Press.
\endbibitem

\bibitem[{O'Brien and Dunson(2004)}]{OBrien_Dunson_2004}
O'Brien, S.~M. and Dunson, D.~B. (2004).
\newblock \enquote{Bayesian Multivariate Logistic Regression.}
\newblock {\em Biometrics\/}, 60(3): 739--746.
\endbibitem

\bibitem[{Polson et~al.(2013)Polson, Scott, and Windle}]{Polson_etal_2013}
Polson, N.~G., Scott, J.~G., and Windle, J. (2013).
\newblock \enquote{Bayesian Inference for Logistic Models Using P{\'o}lya-Gamma
  Latent Variables.}
\newblock {\em Journal of the American Statistical Association\/}, 108(504):
  1339--1349.
\endbibitem

\bibitem[{Rousseeuw and Christmann(2003)}]{Rousseeuw_Christmann_2003}
Rousseeuw, P.~J. and Christmann, A. (2003).
\newblock \enquote{Robustness Against Separation and Outliers in Logistic
  Regression.}
\newblock {\em Computational Statistics and Data Analysis\/}, 42: 315--332.
\endbibitem

\bibitem[{Saban{\'e}s~Bov{\'e} and Held(2011)}]{Bove_Held_2011}
Saban{\'e}s~Bov{\'e}, D. and Held, L. (2011).
\newblock \enquote{Hyper-$g$ Priors for Generalized Linear Models.}
\newblock {\em Bayesian Analysis\/}, 6(3): 387--410.
\endbibitem

\bibitem[{Speckman et~al.(2009)Speckman, Lee, and Sun}]{Speckman_etal_2009}
Speckman, P.~L., Lee, J., and Sun, D. (2009).
\newblock \enquote{Existence of the {MLE} and Propriety of Posteriors for a
  General Multinomial Choice Model.}
\newblock {\em Statistica Sinica\/}, 19: 731--748.
\endbibitem

\bibitem[{Yang and Berger(1996)}]{Yang_Berger_1996}
Yang, R. and Berger, J.~O. (1996).
\newblock \enquote{A Catalog of Noninformative Priors.}
\newblock {\em Institute of Statistics and Decision Sciences, Duke
  University\/}.
\endbibitem

\bibitem[{Zellner and Siow(1980)}]{Zellner_Siow_1980}
Zellner, A. and Siow, A. (1980).
\newblock \enquote{Posterior Odds Ratios for Selected Regression Hypotheses.}
\newblock In {\em Bayesian Statistics: Proceedings of the First International
  Meeting Held in Valencia (Spain)\/}, 585--603. Valencia, Spain: University of
  Valencia Press.
\endbibitem

\bibitem[{Zorn(2005)}]{Zorn_2005}
Zorn, C. (2005).
\newblock \enquote{A Solution to Separation in {B}inary Response Models.}
\newblock {\em Political Analysis\/}, 13(2): 157--170.
\endbibitem

\end{thebibliography}

\end{document}